\documentclass[journal,twoside]{IEEEtran}
\ifCLASSINFOpdf
  \usepackage[pdftex]{graphicx}
\else
  \usepackage[dvips]{graphicx}
\fi
\usepackage{amssymb,amsthm,amsfonts}
\usepackage[cmex10]{amsmath}
  \usepackage[caption=false,font=footnotesize]{subfig}
\usepackage{bm}

\newcommand{\mat}[1]{\ensuremath{\bm{\mathrm{#1}}}}
\newcommand{\A}{\ensuremath{\mat{A}}}
\newcommand{\I}{\ensuremath{\mat{I}}}
\newcommand{\x}{\ensuremath{\mat{x}}}
\newcommand{\thet}{\ensuremath{\mat{\theta}}}
\newcommand{\y}{\ensuremath{\mat{y}}}
\newcommand{\e}{\ensuremath{\mat{e}}}
\newcommand{\Ps}{\ensuremath{\bm{\Psi}}}
\newcommand{\Ph}{\ensuremath{\mat{\Phi}}}

\newcommand{\0}{\ensuremath{\mat{0}}}
\newcommand{\U}{\ensuremath{\mat{U}}}
\newcommand{\Uuk}{\ensuremath{{\underline U}_{\, k}}}
\newcommand{\YuC}{\ensuremath{{\underline Y}_{\, \mathsf{C}}}}
\newcommand{\YuIone}{\ensuremath{{\underline Y}_{\, \mathsf{I},1}}}
\newcommand{\YuItwo}{\ensuremath{{\underline Y}_{\, \mathsf{I},2}}}
\newcommand{\UO}{\ensuremath{\mat{U}_{\Omega}}}
\newcommand{\UpO}{\ensuremath{\mat{U}^\dagger_{\Omega}}}
\newcommand{\UpOT}{\ensuremath{\mat{U}^{\dagger \ T}_{\Omega}}}
\newcommand{\R}{\ensuremath{\mathbb{R}}}
\newcommand{\IR}{\mathsf{IR}}
\newcommand{\JR}{\mathsf{JR}}
\newcommand{\EC}{\mathsf{EC}}
\newcommand{\ECCS}{\mathsf{EC-CS}}
\newcommand{\CS}{\mathsf{CS}}
\newcommand{\var}{\ensuremath{\sigma^2}}
\newcommand{\trasp}[1]{\ensuremath{#1 ^\mathsf{T}}}
\newcommand{\mean}[1]{\ensuremath{\mathbb{E}\left[#1\right]}}
\newcommand{\N}{\mathcal{N}}
\def\df{\triangleq}
\def\Ri{\mathbb{R}}
\newcommand{\lzeronorm}[1]{\ensuremath{\left\| #1\right\|_{0}}}
\newcommand{\lonenorm}[1]{\ensuremath{\left\| #1\right\|_{1}}}
\newcommand{\ltwonorm}[1]{\ensuremath{\left\| #1\right\|_{2}}}
\newtheorem{lemma}{Lemma}
\newtheorem{theorem}[lemma]{Theorem}
\newtheorem{definition}[lemma]{Definition}

\begin{document}
\title{Operational Rate--Distortion Performance of \\ Single--source and Distributed Compressed Sensing}

\author{Giulio~Coluccia,~\IEEEmembership{Member,~IEEE,}
		Aline~Roumy,~\IEEEmembership{Member,~IEEE,}
        and~Enrico~Magli,~\IEEEmembership{Senior Member,~IEEE}% <-this % stops a space
\thanks{G. Coluccia and E. Magli are with Politecnico di Torino, Dipartimento di Elettronica e Telecomunicazioni (email: giulio.coluccia@polito.it, enrico.magli@polito.it). Their research has received funding from the European Research Council under the European Community's Seventh Framework Programme (FP7/2007-2013) / ERC Grant agreement number 279848.
}% <-this % stops a space
\thanks{A. Roumy is with INRIA, campus de Beaulieu, Rennes, France (email: aline.roumy@inria.fr).}% <-this % stops a space
}


\markboth{To appear in IEEE Transactions on Communications}%
{Coluccia \MakeLowercase{\textit{et al.}}: Operational Rate--Distortion Performance of Single--source and Distributed Compressed Sensing}



\maketitle

\begin{abstract}

We consider correlated and distributed sources without cooperation at the encoder.
For these sources, we derive the best achievable performance in the rate-distortion sense of any distributed compressed sensing scheme, under the constraint of high--rate quantization. Moreover, under this model we derive a closed--form expression of the rate gain achieved by taking into account the correlation of the sources at the receiver and a closed--form expression of the average performance of the oracle receiver for independent and joint reconstruction. Finally, we show experimentally that the exploitation of the correlation between the sources performs close to optimal and that the only penalty is due to the missing knowledge of the sparsity support as in (non distributed) compressed sensing. Even if the derivation is performed in the large system regime, where signal and system parameters tend to infinity, numerical results show that the equations match simulations for parameter values of practical interest.
\end{abstract}

% \fi
\IEEEpeerreviewmaketitle

\section{Introduction}\label{sec:intro}

\IEEEPARstart{D}{istributed} sources naturally arise in wireless sensor networks, where sensors may acquire over time  several readings of the same natural quantity, \emph{e.g.}, temperature, in different points of the same environment. Such data must be transmitted to a fusion center for further processing. However, since radio access is the most energy--consuming operation in a wireless sensor network, data transmission among sensors needs to be minimized in order to maximize sensors' battery life. This calls for lossy compression techniques to find a cost--constrained representation in order to exploit data redundancies. In particular, following the example above, sensor readings may vary slowly over time, and hence consecutive readings have similar values, because of the slow variation of the underlying physical phenomenon. Moreover, inter--sensor correlations also exist, as the sensors may be located in the same environment, in which the temperature is rather uniform, leading to compressibility of each single signal and of the set of signals as an ensemble. The question hence arises of how to exploit such correlations in a distributed way, {\em i.e.,} without communication among the sensors, and employing a low--complexity signal representation in order to minimize energy consumption.

In this framework, compressed sensing (CS) \cite{donoho2006cs,candes2006nos} has emerged in past years as an efficient technique for sensing a signal with fewer coefficients than dictated by classic Shannon/Nyquist theory. The hypothesis  underlying this approach is that the signal to be sensed must have a sparse -- or at least compressible -- representation in a convenient basis. In CS, sensing is performed by taking a number of linear projections of
the signal onto pseudorandom sequences. Therefore, the acquisition presents appealing properties. For example, it has low encoding complexity, since no sorting of the sparse signal {coefficients} is required. Moreover,  the choice of the sensing matrix  is blind to the source distribution.  

Using CS as signal representation requires to cast the representation/coding problem in a rate--distortion (RD) framework, particularly regarding the rate necessary to encode the measurements.
 For single sources, this problem has been addressed by several authors. In \cite{dai2011quantized}, a RD analysis of CS reconstruction from quantized measurements was performed, when the observed signal is sparse. Instead, \cite{weidmann12RDsparse} considered the RD behavior of strictly sparse or compressible memoryless sources in their own domain. \cite{fletcher2007rate,goyal2008compressive} considered  the cost of encoding the random measurements for single sources. More precisely, RD analysis was performed and it was shown that adaptive encoding, taking into account the source distribution, outperforms scalar quantization of random measurements at the cost of higher computational complexity. However, in the distributed context, adaptive encoding may loose the inter-correlation between the sources since it is adapted to the distribution of each single source or even the realization of each source. 

On the other hand, the distributed case is more sophisticated. Not only one needs to encode a source, but also to design a scheme capable of exploiting the correlation among different sources. Therefore, distributed CS (DCS) was proposed in \cite{baron2005distributed} and further analyzed in \cite{duarte2013distributed}. In those papers, an architecture for separate acquisition and joint reconstruction was defined, along with three different joint sparsity models (which where merged into a single formulation in the latter paper). For each model, necessary conditions were posed on the number of measurement to be taken on each source to ensure perfect reconstruction. An analogy between DCS and Slepian--Wolf distributed source coding was depicted, in terms of the necessary conditions about the number of measurements, depending on the sparsity degree of sources, and the necessary conditions on encoding rate, depending on conditional and joint entropy, typical of Slepian--Wolf theory. Moreover, it was shown that a distributed system based on CS could save up to 30\% of measurements with respect to separate CS encoding/decoding of each source. On the other hand, \cite{baron2005distributed} extended CS in the acquisition part, but, like CS \cite{donoho2006cs,candes2006nos}, was mainly concerned with the performance of perfect reconstruction, and did not consider the representation/coding problem, which is one of the main issues of a practical scheme and a critical aspect of CS.

In \cite{coluccia2011lossy}, we proposed a DCS scheme that takes into account the encoding cost, exploits both inter- and intra--correlations, and has low complexity. The main idea was to exploit the knowledge of side information (SI) not only as a way to reduce the encoding rate, but also in order to improve the reconstruction quality, as is common in the Wyner--Ziv context \cite{liu2006slepian}. The proposed architecture applies the same Gaussian random matrix to information and SI sources, then quantizes and encodes the measurements with a SW source code.

In this paper, we study analytically the best achievable RD performance of any single--source and distributed CS scheme, under the constraint of high--rate quantization, providing simulation results that perfectly match the theoretical analysis. In particular, we provide the following contributions. First, we derive the asymptotic (in the rate and in the number of measurements) distribution of the measurement vector. Even if the analysis is asymptotic, we show that the convergence to a Gaussian distribution occurs with parameter values of practical interest. Moreover, we provide an analytical expression of the rate gain obtained exploiting inter--source correlation at the decoder. Second, we provide a closed--form expression of the average reconstruction error using the oracle receiver, improving the results existing in literature, consisting only in bounds hardly comparable to the results of numerical simulations \cite{DBLP:journals/corr/abs-1104-4842, laska2012regime}. The proof relies on recent results on random matrix theory \cite{Cook2011}. Third, we provide a closed--form expression of the rate gain due to joint reconstruction from the measurements of multiple sources. We compare the results obtained by theory both with the ideal oracle receiver and with a practical algorithm \cite{coluccia2011lossy}, showing that the penalty with respect to the ideal receiver is due to the lack of knowledge of the sparsity support in the reconstruction algorithm. Despite this penalty, the theoretically--derived rate gain matches that obtained applying  distributed source coding followed by joint reconstruction  to a practical reconstruction scheme. With respect to \cite{baron2005distributed,duarte2013distributed}, we use information theoretic tools to provide an analytical characterization of the performance of CS and DCS, for a given number of measurements and set of system parameters.

This paper is organized as follows. Some background information about source coding with side information at the decoder and CS is given in section~\ref{sec:background}. Novel analytical results are presented in sections~\ref{sec:RD_CS} and \ref{sec:RD_DCS}. These results are validated via numerical simulations that are presented in section~\ref{sec:num_res}. Finally, concluding remarks are given in section~\ref{sec:conclusions}.

\section{Background}\label{sec:background}

\subsection{Notation and definitions}\label{sec:notation}

We denote (column-) vectors and matrices by lowercase and uppercase boldface
characters, respectively. The $(m,n)$-th element of a matrix $\A$ is $(\A)_{m,n}$. The $m$-th row of matrix $\A$ is $(\A)_m$. The $n$-th element of column vector $\mat{v}$ is $(\mat{v})_n$. The transpose of a matrix $\A$ is $\trasp{\A}$. %The stack operator $\vect{\A}$ denotes the column vector obtained by stacking the columns of \A\ on top of each other, from left to right.

The notation $\lzeronorm{\mat{v}}$ denotes the number of nonzero elements of vector
$\mat{v}$. The notation $\lonenorm{\mat{v}}$ denotes the $\ell_1$-norm of the vector $\mat{v}$
and is defined as $\lonenorm{\mat{v}} \df \sum_i \left |(\mat{v})_i\right |$~. The
notation $\ltwonorm{\mat{v}}$ denotes the Euclidean norm of the vector $\mat{v}$ and is
defined as $\ltwonorm{\mat{v}} \df \sqrt{\sum_i \left |(\mat{v})_i\right |^2}$~. The
notation $A\sim\N(\mu,\sigma^2)$ means that the random variable $A$ is Gaussian
distributed, its mean is $\mu$, and its variance is $\sigma^2$. Additional notation will be defined throughout the paper where appropriate.

\subsection{Source Coding with Side Information at the decoder}\label{sec:dsc}
Source Coding with SI at the decoder refers to the problem of compressing a source $X$ when another source $Y$, correlated to $X$, is available at the decoder only. It is a special case of distributed source coding, where the two sources have to be compressed without any cooperation at the encoder. 

For lossless compression, if $X$ is compressed without knowledge of $Y$  at its conditional entropy, \emph{i.e.}, $R_X > H(X|Y)$, it can be recovered with vanishing error rate exploiting $Y$ as SI. This represents the asymmetric setup, where  source $Y$ is compressed in a lossless way ($R_Y>H(Y)$) or otherwise known at the decoder. Therefore, the lack of SI at the encoder does not incur any compression loss with respect to joint encoding, as the total rate required by DSC is equal to $H(Y)+H(X|Y)=H(X,Y)$. The result holds for i.i.d. finite sources $X$ and $Y$ \cite{slepian1973noiseless} but also for ergodic discrete sources \cite{cover1975aproof}, or when $X$ is i.i.d. finite, $Y$ is i.i.d. continuous and is available at the decoder \cite[Proposition~19]{rebollo2006highrate}.

For lossy compression of i.i.d sources, \cite{wyner1978continuousiid} shows that the lack of SI at the encoder incurs a loss except for some distributions (Gaussian sources, or more generally Gaussian correlation noise). Interestingly, \cite{rebollo2006highrate} shows that uniform scalar quantization followed by lossless compression incurs a suboptimality of $1.53$~dB, in the high--rate regime. Therefore, practical solutions (see for example \cite{bassi2010wyner}) compress and decompress the data relying on an \textit{inner lossless distributed codec}, usually referred to as Slepian--Wolf Code (SWC), and an \textit{outer quantization--plus--reconstruction filter}.

\subsection{Compressed Sensing}\label{sec:CS}

In the standard CS framework, introduced in \cite{donoho2006cs,candes2006nos}, a signal $\x\in\Ri^{N\times 1}$
 which has  a sparse representation in some basis $\Ps\in\Ri^{N\times N}$, \textit{i.e.}:
\begin{equation*}
\x = \Ps \bm{\theta},\quad \lzeronorm{\bm{\theta}} = K,\quad K\ll N
\end{equation*}
can be recovered by a smaller vector of linear measurements $\y = \Ph\x$, $\y\in\Ri^{M\times 1}$ and $K<M<N$,  where $\Ph\in\Ri^{M\times N}$ is the \emph{sensing matrix}. The optimum solution, requiring at least $M = 2K $ measurements, would be
$$
\widehat{\bm{\theta}}=\arg\min_{\bm{\theta}}\lzeronorm{\bm{\theta}}\ \quad \text{s.t.}\quad \Ph\Ps\bm{\theta} = \y~.
$$
Since the $\ell_0$ norm minimization is an NP-hard problem,
one can resort to a linear programming reconstruction by minimizing
the $\ell_1$ norm
\begin{equation}\label{eq:CS_recovery}
\widehat{\bm{\theta}}=\arg\min_{\bm{\theta}}\lonenorm{\bm{\theta}}\ \quad \text{s.t.}\quad \Ph\Ps\bm{\theta} = \y~
\end{equation}
and $\widehat{\x}=\Ps\widehat{\bm{\theta}}$, provided that $M = O(K\log(N/K))$ \cite{candes2006nos}.

When the measurements are noisy, i.e. when $\y = \Ph\x + \mat{e}$, where $\mat{e}\in\Ri^{M\times 1}$ is the vector representing additive noise such that $\ltwonorm{\mat{e}} < \varepsilon$, $\ell_1$ minimization
with inequality constraints is used for reconstruction:
\begin{equation}\label{eq:CS_recovery_relaxed}
\widehat{\bm{\theta}}=\arg\min_{\bm{\theta}}\lonenorm{\bm{\theta}}\ \quad \text{s.t.}\quad \ltwonorm{\Ph\Ps\bm{\theta} - \y} < \varepsilon~
\end{equation}
and $\widehat{\x}=\Ps\widehat{\bm{\theta}}$, known as Basis Pursuit DeNoising \textsf{(BPDN)}, provided that $M = O(K\log(N/K))$ and that each submatrix consisting of $K$ columns of $\Ph\Ps$  is (almost) distance preserving \cite[Definition 1.3]{eldar2012compressed}. The latter condition is the \emph{Restricted Isometry Property} (RIP). Formally, the matrix $\Ph\Ps$ satisfies the RIP of order $K$ if $\exists \delta_K \in (0,1]$ such that, for any $\bm{\theta}$ with $\lzeronorm{\bm{\theta}} \le K$:
\begin{equation}
(1-\delta_K)\ltwonorm{\bm{\theta}}^2\le\ltwonorm{\Ph\Ps\bm{\theta}}^2\le(1+\delta_K)\ltwonorm{\bm{\theta}}^2,
	\label{eq:RIP}
\end{equation}
where $\delta_K$ is the RIP constant of order $K$. It has been shown in \cite{baraniuk2008spr} that when $\Ph$ is an i.i.d. random matrix drawn from any subgaussian distribution and $\Ps$ is an orthogonal matrix, $\Ph\Ps$ satisfies the RIP with overwhelming probability.

\section{Rate--Distortion functions of\\ Single--source Compressed Sensing}\label{sec:RD_CS}

In this section, we derive the best achievable performance in the RD sense over all CS schemes, under the constraint of high--rate quantization. The novel result about the distribution of the measurements derived in Theorem~\ref{theorem:y iid gaussian} allows to write a closed--form expression of the RD functions of the measurement vectors. In Theorem~\ref{th:RD reconstruction non distributed}, we derive a novel closed--form expression of the average reconstruction error of the oracle receiver, which will use the results from Theorem~\ref{theorem:y iid gaussian} to present the RD functions of the reconstruction.

\subsection{System Model}

\begin{definition}\label{def:sparse} 
\emph{(Sparse vector).}
The vector $\x \in \R^N$ is said to be $(K,N,\var_{\theta},\Ps)$-\emph{sparse} if
$\x$ is sparse in the domain defined by
the orthogonal matrix $\Ps\in\Ri^{N\times N}$, namely:
$\x = \Ps\bm{\theta}$, with $\lzeronorm{\bm{\theta}} = K$, and if the nonzero components of $\thet$ are modeled as i.i.d. centered random variables with variance  $\var_{\theta}<\infty$. \Ps\ is independent of $\bm{\theta}$.
\end{definition}
The sparse $\x$ vector is observed through a smaller vector of Gaussian measurements defined as
\begin{definition}\label{def:measurement} 
\emph{(Gaussian measurement).}
The vector $\y$ is called the $(M,N,\var_{\Phi},\Ph)$-\emph{Gaussian measurement} of $\x\in\Ri^{N}$, if $\y = \frac{1}{\sqrt{M}} \Ph\x $, where the sensing matrix $\Ph\in\Ri^{M \times N}$, with $M<N$, is a random matrix\footnote{This definition complies with the usual form $\y = \Ph\x $ where the variance $\var_{\Phi}$ of the elements of \Ph\ depends on $M$. Here, we wanted to keep $\var_{\Phi}$ independent of system parameters.} with i.i.d. entries drawn from $\N(0,\var_{\Phi})$ with $\var_{\Phi}<\infty$.
\end{definition}
We denote as $\y_{q}$ the quantized version of $\y$. To analyze the RD tradeoff, we consider the large system regime defined below.
\begin{definition}\label{def:asymptotic mode CS} 
\emph{(Large system regime, overmeasuring and sparsity rates).}
Let $\x$ be $(K,N,\var_{\theta},\Ps)$-{sparse}. Let $\y$ be the $(M,N,\var_{\Phi},\Ph)$-Gaussian measurement of $\x$.
The system is said to be  in the \emph{large system regime} if 
$N$ goes to infinity, $K$ and $M$ are functions of $N$ and tend to infinity as $N$ does, under the constraint that the rates $K/N$ and $M/K$ converge to constants called \emph{sparsity rate} ($\gamma$) and \emph{overmeasuring rate}  ($\mu$) i.e.:
\begin{equation}	 \lim\limits_{N \to +\infty} \frac{K}{N}=\gamma,\ \  \lim\limits_{N \to +\infty}\frac{M}{K}= \mu>1
\end{equation}
\end{definition}

The sparsity rate is a property of the signal. Instead, the overmeasuring rate is the ratio of the number of measurements to the number of non zero components and is therefore a property of the system \cite{dai2011quantized}.

\subsection{Rate--distortion functions of measurement vector}
%====================================================================
\label{subsec:RD_CS measurement}

The \emph{information RD function} of an i.i.d. source $X$ defines the minimum amount of information per source symbol $R$ needed to describe the source under the distortion constraint $D$. 
For an i.i.d. Gaussian source $X \sim \N(0,\var_x)$, choosing as distortion metric the squared error between the source and its representation on $R$ bits per symbol, the distortion satisfies
\begin{equation}
	D_x(R) = \var_x 2^{-2R}.
	\label{eq:RD X}
\end{equation}
Interestingly, the \emph{operational RD function} of the same Gaussian source, with uniform scalar quantizer and entropy coding satisifies, in the high--rate regime: 
\begin{equation}
\lim_{R\to +\infty} \frac{1}{\var_x} 2^{2R} D_x^{\EC}(R) = \frac{\pi e}{6}
	\label{eq:RD ec X}
\end{equation}
where $\EC$ stands for \textit{entropy--constrained scalar quantizer}. This leads to a $1.53$~dB gap between the information and the operational RD curves. \eqref{eq:RD ec X} can be easily extended to other types of quantization adapting the factor $\frac{\pi e}{6}$ to the specific quantization scheme.

\begin{theorem}
\label{theorem:y iid gaussian}
\emph{(CS: Asymptotic distribution of Gaussian measurements and measurement RD function).}
Let $\x$ be $(K,N,\var_{\theta},\Ps)$-{sparse}. Let $\y$ be the $(M,N,\var_{\Phi},\Ph)$-Gaussian measurement of $\x$, s.t. $K<M<N$. Consider the large system regime with finite sparsity rate $\gamma=\lim_{N\to \infty}\frac{K}{N}$ and finite overmeasuring rate $\mu=\lim_{N\to \infty}\frac{M}{K}>1$.\\ The Gaussian measurement converges in distribution to an i.i.d. Gaussian, centered random sequence with variance 
\begin{equation}
	\sigma_y^2 = \frac{1}{\mu} \ \var_{\Phi} \var_{\theta}.
\end{equation}
Therefore, the information RD function satisifies 
\begin{align}
& \lim_{N \to +\infty}
\frac{1}{\var_{y}} 2^{2R} D_{y}(R) = 1,
\end{align}
where $R$ is the encoding rate per {measurement} sample,
and the entropy--constrained scalar quantizer achieves a distortion $D^{\EC}_{y_1}$ that satisfies
\begin{align}
&\lim_{R  \to +\infty} \ \lim_{N \to +\infty}
\frac{1}{\var_{y}} 2^{2R} D^{\EC}_{y}(R) = \frac{\pi e}{6}~.
\label{eq:rd_ec_y}
\end{align}
\end{theorem}

\noindent \textbf{{Sketch of proof.}} The distribution of the Gaussian matrix $\Ph$ is invariant under orthogonal transformation. Thus, we obtain $\y=\frac{1}{\sqrt{M}} \Ph\Ps\thet=\frac{1}{\sqrt{M}}\mat{U}\thet$, where $\mat{U}$ is an i.i.d. Gaussian matrix with variance $\var_{\Phi}$. Then, we consider a finite length subvector $\y^{(m)}$ of $\y$. From the multidimensional Central Limit theorem (CLT) \cite[Theorem 7.18]{khoshnevisan2007}, $\y^{(m)}$ converges to a Gaussian centered vector with independent components. Then, as $m\to \infty$, the sequence of Gaussian measurements converges to an i.i.d. Gaussian sequence. See Appendix~\ref{annex:lemma y iid gaussian} for the complete proof.
\hfill $\square$

Theorem~\ref{theorem:y iid gaussian} generalizes \cite[Theorem 2]{dai2011quantized}, which derives the marginal distribution of the measurements,  when the observed signal is directly sparse. Instead, we derive the \textit{joint} distribution of the measurements and consider \textit{transformed} sparse signals. 

We stress the fact that, even if the RD curves for measurement vectors do not have any ``practical'' direct use, they are required to derive the RD curves for the reconstruction of the sources, which can be found later in this section.

\subsection{Rate--distortion functions of the reconstruction}
% ===============================================================
\label{subsec:RD_CS reconstruction}

We now evaluate the performance of CS reconstruction with quantized  measurements. The performance depends on the amount of noise affecting the measurements. In particular, the distortion ${\ltwonorm{\widehat{\x}-\x}^2}$ is upper bounded by the
noise variance up to a scaling factor \cite{candes2006ssr, candes2008restricted}.
\begin{equation}\label{eq:rec_upper_bound}
\ltwonorm{\widehat{\x}-\x}^2 \le c^2\varepsilon^2~,
\end{equation}
where the constant $c$  depends on the realization of the measurement matrix, since it is a function of the RIP constant. Since we consider the average\footnote{The average performance is obtained averaging over all random variables i.e. the measurement matrix, the non-zero components $\thet$ and noise, as for example in \cite{laska2012regime}.} performance, we need to consider the worst case $c$ and this upper bound will be very loose \cite[Theorem 1.9]{eldar2012compressed}.

Here, we consider the \emph{oracle} estimator, which is the estimator knowing exactly the sparsity support $\Omega=\{i|\bm{\theta}_i\neq 0\}$ of the signal $\x$. For the oracle estimator, upper and lower bounds depending on the RIP constant can be found, for example in \cite{DBLP:journals/corr/abs-1104-4842} when the noise affecting the measurements is white and in \cite{laska2012regime} when the noise is correlated. Unlike \cite{DBLP:journals/corr/abs-1104-4842, laska2012regime}, in this paper the average performance of the oracle, depending on system parameters only, is derived exactly.

As we will show in the following sections, the characterization of the ideal oracle estimator allows to derive the reconstruction RD functions with results holding also when non ideal estimators are used.

\begin{theorem} 
\label{th:RD reconstruction non distributed}
\emph{(CS: Reconstruction RD functions).}
Let $\x$ be $(K,N,\var_{\theta},\Ps)$-{sparse}. Let $\y$ be the $(M,N,\var_{\Phi},\Ph)$-Gaussian measurement of $\x$, s.t. $K+3<M<N$. Consider the large system regime with finite sparsity rate $\gamma=\lim_{N\to \infty}\frac{K}{N}$ and finite overmeasuring rate $\mu=\lim_{N\to \infty}\frac{M}{K}>1$. $R$ denotes the encoding rate per {measurement} sample.\\
Assume reconstruction by the oracle estimator, when the support $\Omega$ of $\x$ is available at the receiver. The operational RD function of any CS reconstruction algorithm is lower bounded by that of the oracle estimator that satisfies 
\begin{equation}
D^{\CS}_{x}(R) \ge D^{\text{\emph{oracle}}}_{x}(R) =\gamma\frac{\mu}{\mu-1} \frac{1}{\sigma^2_\Phi} D_{y}(R)= \frac{\gamma}{\mu-1} \sigma^2_\theta  2^{-2R}.
\label{eq:RD recons x}
\end{equation}
Similarly, the entropy-constrained RD function satisfies in the high-rate regime 
\begin{equation}
D^{\ECCS}_{x}(R) \ge D^{\EC \ \text{\emph{oracle}}}_{x}(R) =  \frac{\gamma}{\mu-1} \sigma^2_\theta  \frac{\pi e}{6} 2^{-2R}.
\label{eq:EC RD recons x}
\end{equation}
\end{theorem}

\noindent \textbf{{Sketch of proof.}} We use a novel result about the expected value of a matrix following a generalized inverse Wishart distribution \cite[Theorem 2.1]{Cook2011}. This result can be applied to the distortion of the oracle estimator for finite--length signals, depending on the expected value of the pseudo inverse of Wishart matrix \cite{DiazGarcia2006}. The key consequence is that the distortion of the oracle only depends on the variance of the quantization noise and not on its covariance matrix. Therefore, our result holds even if the noise is correlated (for instance if vector quantization is used). Hence, this result applies to any quantization algorithm. This result improves those in \cite[Theorem 4.1]{DBLP:journals/corr/abs-1104-4842} and \cite{laska2012regime},  where upper and lower bounds depending on the RIP constant of the sensing matrix are given, and it also generalizes \cite[section III.C]{dai2011quantized}, where a lower bound is derived whereas we derive the exact average performance. See Appendix~\ref{annex:th:RD reconstruction non distributed} for the complete proof.
\hfill $\square$

It must be noticed that the condition $M>K+3$ is not restrictive since in all cases of practical interest, $M>2K$.

\section{Rate--Distortion functions of\\ Distributed Compressed Sensing}\label{sec:RD_DCS}

\begin{figure*}
\centering
\includegraphics[width=\textwidth]{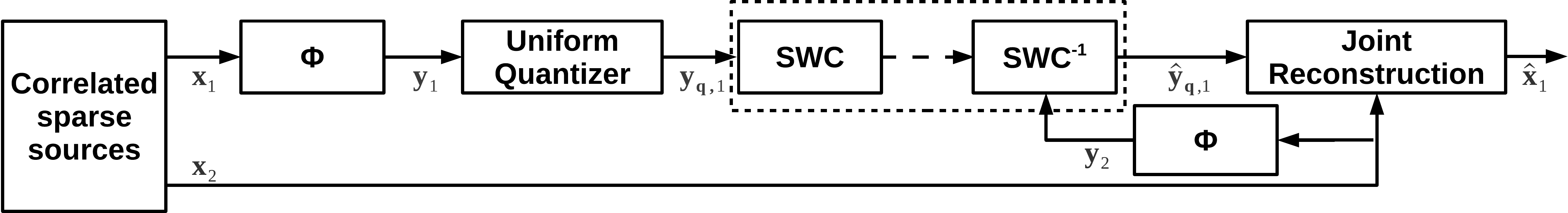}
\caption{The Distributed Compressed Sensing scheme in \cite{coluccia2011lossy}}
\label{fig:bd_mq}
\end{figure*}

In this section, we derive the best achievable performance in the RD sense over all DCS schemes, under the constraint of high--rate quantization. Note that \cite{coluccia2011lossy} (see Fig.~\ref{fig:bd_mq}) is one instance of such a scheme. Novel results about the distribution of the measurements in the distributed case are presented in Theorem~\ref{prop:rd_dcs}. Hence, Theorem~\ref{th:RD reconstruction}, will combine the results of Theorem~\ref{prop:rd_dcs} and previous section to derive the RD functions of the reconstruction in the distributed case.

\subsection{Distributed System Model}

\begin{definition}\label{def:jsm1} 
\emph{(Correlated Sparse vectors).}\\
$J$ vectors $\x_j\in \Ri^{N \times 1}, j \in\{1,\ldots,J\}$ are said to be $(\{K_{\mathsf{I},j}\}_{j=1}^J,K_{\mathsf{C}},\{K_{j}\}_{j=1}^J,N,\var_{\theta_\mathsf{C}},\{\var_{\theta_{\mathsf{I},j}}\}_{j=1}^J,\Ps)$ -\emph{sparse} if: \\
\textbf{\emph{i)}} Each vector $\x_j = \x_\mathsf{C} +
\x_{\mathsf{I},j}$ is the sum of a \emph{common} component $\x_\mathsf{C}$ shared by all signals and  an \emph{innovation} component $\x_{\mathsf{I},j}$, which is unique to each signal $\x_j$.\\
\textbf{\emph{ii)}} Both $\x_\mathsf{C}$ and $\x_{\mathsf{I},j}$ are sparse in the same domain defined by
the orthogonal matrix $\Ps\in\Ri^{N\times N}$, namely:
$\x_\mathsf{C} = \Ps\bm{\theta}_\mathsf{C}$ and $\x_{\mathsf{I},j} =
\Ps\bm{\theta}_{\mathsf{I},j}$, with
$\lzeronorm{\bm{\theta}_\mathsf{C}} = K_\mathsf{C}$,
$\lzeronorm{\bm{\theta}_{\mathsf{I},j}} = K_{\mathsf{I},j}$ and $ K_\mathsf{C},
K_{\mathsf{I},j} < N$.\\ 
\textbf{\emph{iii)}} The global sparsity of $\x_j$ is $K_{j}$, with $\max\left\lbrace K_\mathsf{C},K_{\mathsf{I},j}\right\rbrace\le K_{j}\le K_\mathsf{C}+K_{\mathsf{I},j}$.\\
\textbf{\emph{iv)}} The nonzero components of $\thet_\mathsf{C}$ and $\thet_{\mathsf{I},j}$ are i.i.d. centered random variables with variance  $\var_{\theta_\mathsf{C}}<\infty$ and $\var_{\theta_{\mathsf{I},j}}<\infty$, respectively.
\end{definition}
The correlation between the sources is modeled through a \emph{common} component and their difference through an individual \emph{innovation} component. This is a good fit for signals acquired by a group of sensors monitoring the same physical event in different spacial positions, where local factors can affect the innovation component of a more global behavior taken into account by this common component. Note that the Joint Sparsity Model-1 (JSM-1) \cite{baron2005distributed} and the ensemble sparsity model (ESM) in \cite{duarte2013distributed} are deterministic models. Instead, the sparse model (Def.~\ref{def:jsm1}) is probabilistic, since we look for the performance averaged over all possible realizations of the sources.

Focusing without loss of generality on the case $J=2$, we assume that $\x_1$ and $\x_2$ are $(K_{\mathsf{I},1},K_{\mathsf{I},2},K_{\mathsf{C}},K_1,K_2,N,$ $ \var_{\theta_\mathsf{C}},\var_{\theta_{\mathsf{I},1}},\var_{\theta_{\mathsf{I},2}},\Ps )$-{sparse}.
$\x_1$ is the source to be compressed whereas $\x_2$ serves as SI. $\y_1$ and $\y_2$ are the $(M,N,\var_{\Phi},\Ph)$-Gaussian measurements of $\x_1$ and $\x_2$, and $\y_{q,j}$ is the quantized version of $\y_j$.

The large system regime becomes in the distributed case:
\begin{definition}\label{def:asymptotic model} 
\emph{(Large system regime, sparsity andovermeasuring rates, and overlaps).}\\
Let the $J$ vectors $\x_j\in \Ri^{N \times 1},$ $\forall j \in\{1,\ldots,J\}$ be $(\{K_{\mathsf{I},j}\}_{j=1}^J,K_{\mathsf{C}},\{K_{j}\}_{j=1}^J,N,\var_{\theta_\mathsf{C}},\{\var_{\theta_{\mathsf{I},j}}\}_{j=1}^J,\Ps)$ -sparse. For each $j$, let $\y_j\in \Ri^{M \times 1}$ be the $(M,N,\var_{\Phi},\Ph)$-\emph{Gaussian measurement} of $\x_j$.
The system is said to be  in the \emph{large system regime} if: \\
\textbf{\emph{i)}}
$N$ goes to infinity.\\
\textbf{\emph{ii)}}
The other dimensions $K_{\mathsf{I},j} ,K_{\mathsf{C}},K_{j} , M$ are functions of $ N$ and tend to infinity as $N$ does.\\
\textbf{\emph{iii)}} The following rate converges to a constant called \emph{sparsity rate} as $N$ goes to infinity:
\begin{equation}
	\frac{K_j}{N} \to \gamma_j~. 
\end{equation}
\textbf{\emph{iv)}} The following rate converges to a constant called \emph{overmeasuring rate} as $N$ goes to infinity:
\begin{equation}
\frac{M}{K_j} \to \mu_j>1.
\end{equation}
\textbf{\emph{v)}}  All following rates converge to constants called \emph{overlaps} of the common and innovation components as $N$ goes to infinity:
\begin{equation}
\frac{K_{\mathsf{C}}}{K_j} \to \omega_{\mathsf{C},j}, \ \frac{K_{\mathsf{I},j}}{K_j} \to \omega_{\mathsf{I},j}~.
\end{equation}
Note that $\max\left\lbrace\omega_{C,j} , \omega_{I,j}\right\rbrace \le 1 \le \omega_{C,j} + \omega_{I,j} \le 2$.
\end{definition}

\subsection{Rate--distortion functions of measurement vector}
%====================================================================
\label{subsec:distribCS RD measurement}

The information RD function can also be derived for a pair $(X,Y) \sim \N(0,\mat{K}_{xy})$ of i.i.d. jointly Gaussian distributed random variables with covariance matrix
\begin{equation}
\mat{K}_{xy} = \left(\begin{array}{cc}
\var_x & \rho_{xy} \sigma_x\sigma_y \\
\rho_{xy} \sigma_x\sigma_y & \var_y
\end{array}\right).
	\label{eq:inter-correlation}
\end{equation}
Interestingly, when the SI is available at both encoder and decoder or at the decoder only, the information RD function is the same:
\begin{equation}
%$$
D_{x|y}(R) = \var_x(1-\rho_{xy}^2)2^{-2R} = D_x(R+R^*),
%$$
\label{eq:RD_WZ_def}
\end{equation}
where $R^* = \frac{1}{2}\log_2\frac{1}{1-\rho_{xy}^2}\ge0$ is the \emph{rate gain}, measuring the amount of rate we save by using the side information $Y$ to decode $X$. This result holds for optimal vector quantizer \cite{wyner1978continuousiid} but also for scalar uniform quantizers \cite[Theorem 8 and Corollary 9]{rebollo2006highrate} by replacing $D_x$ in \eqref{eq:RD_WZ_def} by the entropy constrained distortion function $D_x^{\EC}(R)$, defined in \eqref{eq:RD ec X}.

To derive the RD curves for the reconstruction of the sources, we first generalize Theorem~\ref{theorem:y iid gaussian} and derive the asymptotic distribution of pairs of measurements.
\begin{theorem}\label{prop:rd_dcs}
\emph{(Distributed CS: Asymptotic distribution of the pair of Gaussian measurements and measurement RD functions).}
Let $\x_1$ and $\x_2$ be $(K_{\mathsf{I},1},K_{\mathsf{I},2},K_{\mathsf{C}},K_1,K_2,N, \var_{\theta_\mathsf{C}},\var_{\theta_{\mathsf{I},1}},$ $\var_{\theta_{\mathsf{I},2}},\Ps )$-{sparse}. $\x_2$ serves as SI for $\x_1$ and is available at the decoder, only. Let $\y_1$ and $\y_2$ be the $(M,N,\var_{\Phi},\Ph)$-Gaussian measurements of  $\x_1$ and $\x_2$. 
Let $(Y_{1},Y_{2})$ be the pair of random processes associated to the random vectors $(\y_1,\y_2)$. In the large system regime,
$(Y_{1},Y_{2})$ converges to an i.i.d. Gaussian sequence with covariance matrix
\begin{align}
\mat{K}_{12} &= \left(\begin{array}{cc}
\var_{y_1} & \rho_{12} \sigma_{y_1}\sigma_{y_2} \\
\rho_{12} \sigma_{y_1}\sigma_{y_2} & \var_{y_2}
\end{array}\right),
	\label{eq:inter-correlation-matrix12} \\
\sigma^2_{y_j} &= \frac{\var_{\Phi}}{\mu_j} \left[  \omega_{\mathsf{C},j} \var_{\theta_\mathsf{C}} +  \omega_{\mathsf{I},j} \var_{\theta_{\mathsf{I},j}} \right]
\label{eq:var_y1}\\
 \rho_{12} &= \left[
\Big(1 + \frac{ \omega_{\mathsf{I},1}}{ \omega_{\mathsf{C},1}}\frac{\var_{\theta_{\mathsf{I},1}}}{\var_{\theta_\mathsf{C}}}\Big)
\Big(1 + \frac{ \omega_{\mathsf{I},2}}{ \omega_{\mathsf{C},2}}\frac{\var_{\theta_{\mathsf{I},2}}}{\var_{\theta_\mathsf{C}}}\Big)
\right]^{-\frac{1}{2}}.
\label{eq:rho_y1y2} 
\end{align}

Let $R$ be the encoding rate per {measurement} sample.
When the SI is not used, the information RD function satisifies 
\begin{align}
& \lim_{N \to +\infty}
\frac{1}{\var_{y_1}} 2^{2R} D_{y_1}(R) = 1,
\end{align}
and the entropy--constrained scalar quantizer achieves a distortion $D^{\EC}_{y_1}$ that satisfies
\begin{align}
&\lim_{R  \to +\infty} \ \lim_{N \to +\infty}
\frac{1}{\var_{y_1}} 2^{2R} D^{\EC}_{y_1}(R) = \frac{\pi e}{6}~.
\label{eq:rd_ec_y1}
\end{align}

When the measurement $\y_2$ of the SI is used at the decoder, the information RD function satisifies
\begin{equation}
\lim_{N \to +\infty}
\frac{1}{\var_{y_1}} 2^{2(R+R^*)} D_{y_1|y_2}(R) = 1~,
\label{eq:rd_y1y2}
\end{equation}
while the entropy--constrained scalar quantizer achieves a distortion $D^{\EC}_{y_1|y_2}$ that satisfies
\begin{equation}
\lim_{R  \to +\infty} \ \lim_{N \to +\infty}
\frac{1}{\var_{y_1}} 2^{2(R+R^*)} D^{\EC}_{y_1|y_2}(R) = \frac{\pi e}{6}~,
\label{eq:rd_ec_y1y2}
\end{equation}
where
\begin{equation}
R^* = \frac{1}{2}\log_2\frac{1}{1-\rho_{12}^2}~.
\label{eq:rate_gain}
\end{equation}

Therefore, in the large system regime (and in the high--rate regime for entropy constrained scalar quantizer), the measurement $\y_2$ of the SI helps reducing the rate by $R^*$ \eqref{eq:rate_gain} bits per measurement sample:
\begin{equation}
 D_{y_1|y_2}(R) = D_{y_1}(R+R^*).
\label{eq:RD y with rate_gain}
\end{equation}
\end{theorem}

\noindent \textbf{{Sketch of proof.}} We consider a vector of finite length $2m$, which contains the first $m$ components of $\y_1$ followed by the first $m$ components of $\y_2$. The vector can be seen as a sum of three components, where each component converges to a Gaussian vector from the multidimensional CLT \cite[Theorem 7.18]{khoshnevisan2007}. Finally, we obtain that $(Y_{1},Y_{2})$ converges to an i.i.d. Gaussian process. Therefore, classical RD results for i.i.d. Gaussian sources apply. See Appendix~\ref{annex:prop:rd_dcs} for the complete proof.\hfill $\square$

Theorem~\ref{prop:rd_dcs} first states that the measurements of two sparse vectors converge to an i.i.d. Gaussian process in the large system regime. Then, lossy compression of the measurements is considered and the information and entropy constrained rate distortion functions are derived. It is shown that if one measurement vector is used as side information at the decoder, some rate can be saved, depending on the sparse source characteristics, only (see \eqref{eq:rate_gain} and \eqref{eq:rho_y1y2}).

\subsection{Rate--distortion functions of the reconstruction}
% ===============================================================
\label{subsec:RD reconstruction}

We now derive the RD functions after reconstruction of the DCS scheme.
\begin{theorem} 
\label{th:RD reconstruction}
\emph{(Distributed CS: Reconstruction RD functions).} 
Let $\x_1$ and $\x_2$ be 
$(K_{\mathsf{I},1}$ ,$K_{\mathsf{I},2}$,$K_{\mathsf{C}}$,$K_1$,$K_2$,$N$, $\var_{\theta_\mathsf{C}}$,$\var_{\theta_{\mathsf{I},1}}$,$\var_{\theta_{\mathsf{I},2}}$,$\Ps )$-{sparse}. $\x_2$ serves as SI for $\x_1$ and is available at the decoder, only.
Let $\y_1$ and $\y_2$ be the $(M,N,\var_{\Phi},\Ph)$-Gaussian measurements of  $\x_1$ and $\x_2$, s.t. $K_1+3<M<N$. Let $R$ be the encoding rate per {measurement} sample. 
The distortion\footnote{All the RD functions are operational referred to CS reconstruction algorithms, so we omit the \textsf{CS} superscript not to overload the notation} of the source $\x_1$  is denoted as  $D^{\IR}_{{x}_1}$ when the SI is not available at the receiver, $D^{\IR}_{x_1|y_2}$ when the measurements of the SI are available at the SWC decoder ($\IR$ stands for \emph{independent reconstruction}), and $ D^{\JR}_{{x}_1|x_2}$ when the SI is used not only to reduce the encoding rate but also to improve the reconstruction fidelity ($\JR$ stands for \emph{joint reconstruction}).
\\
Then, when independent reconstruction is performed, the RD functions for $\x_1$ satisfy, in the large system regime:
\begin{align}
D^{\IR}_{{x}_1}(R) &\ge D^{\IR \ \emph{oracle}}_{{x}_1}(R) = \gamma_1\frac{\mu_1}{\mu_1-1} \frac{1}{\sigma^2_\Phi} D_{y_1}(R),
	\label{eq:perf_oracle_uncond} \\
D^{\IR}_{x_1|y_2}(R) &\ge D^{\IR\ \emph{oracle}}_{x_1|y_2}(R) = \gamma_1\frac{\mu_1}{\mu_1-1} \frac{1}{\sigma^2_\Phi} D_{y_1|y_2}(R),
	\label{eq:perf_oracle_dsc_IR}
\end{align} 
Therefore,  in the large system regime, the operational RD functions satisfy
\begin{align}
D^{\IR}_{{x}_1}(R) &\ge D^{\IR \ \emph{oracle}}_{{x}_1}(R) =  \gamma_1 \frac{  \omega_{\mathsf{C},1} \var_{\theta_\mathsf{C}} +  \omega_{\mathsf{I},1} \var_{\theta_{\mathsf{I},1}} }{\mu_1-1}   2^{-2R},
	\label{eq:perf_oracle_uncond gaussian} \\
D^{\IR}_{x_1|y_2}(R) &\ge D^{\IR\ \emph{oracle}}_{x_1|y_2}(R) = \gamma_1 \frac{  \omega_{\mathsf{C},1} \var_{\theta_\mathsf{C}} +  \omega_{\mathsf{I},1} \var_{\theta_{\mathsf{I},1}} }{\mu_1-1}  2^{-2(R+R^*)},
	\label{eq:perf_oracle_dsc_IR gaussian}\\
D^{\IR}_{x_1|y_2}(R) &=D^{\IR}_{{x}_1}(R+R^*),
	\label{eq:RD for IR with rate gain gaussian}
\end{align} 
where $R^*$ is defined in  \eqref{eq:rate_gain}. In the large system regime and in the high--rate regime, the entropy constrained RD functions satisfy:
\begin{align}
D^{\IR \ \EC}_{{x}_1}(R) &\ge  \gamma_1 \frac{  \omega_{\mathsf{C},1} \var_{\theta_\mathsf{C}} +  \omega_{\mathsf{I},1} \var_{\theta_{\mathsf{I},1}} }{\mu_1-1}  \frac{\pi e}{6}2^{-2R},
	\label{eq:perf_oracle_uncond gaussian EC} \\
D^{\IR\ \EC}_{x_1|y_2}(R) &\ge \gamma_1\frac{  \omega_{\mathsf{C},1} \var_{\theta_\mathsf{C}} +  \omega_{\mathsf{I},1} \var_{\theta_{\mathsf{I},1}} }{\mu_1-1}  \frac{\pi e}{6} 2^{-2(R+R^*)}.
	\label{eq:perf_oracle_dsc_IR gaussian EC}
\end{align} 

When joint reconstruction is performed, the RD functions for $\x_1$ satisfy:
\begin{align}
D^{\JR}_{{x}_1|x_2}(R) &\ge D^{\JR \ \emph{oracle}}_{{x}_1|x_2}(R) = \omega_{\mathsf{I},1}\gamma_1\frac{\mu_1}{\mu_1 - \omega_{\mathsf{I},1}} \frac{1}{\sigma^2_\Phi}  D_{y_1|y_2}(R),
	\label{eq:perf_oracle_dsc_JR}
\end{align} 
where, in the large system regime,
\begin{align}
& D^{\JR \ \emph{oracle}}_{{x}_1|x_2}(R) =\omega_{\mathsf{I},1}\gamma_1 \frac{  \omega_{\mathsf{C},1} \var_{\theta_\mathsf{C}} +  \omega_{\mathsf{I},1} \var_{\theta_{\mathsf{I},1}} }{\mu_1-\omega_{\mathsf{I},1}}   2^{-2(R+R^*)},
	\label{eq:perf_oracle_dsc_JR gaussian}
\end{align} 
and in the high rate regime
\begin{align}
& D^{\JR \ \EC\ \emph{oracle}}_{{x}_1|x_2}(R) =\omega_{\mathsf{I},1}\gamma_1  \frac{  \omega_{\mathsf{C},1} \var_{\theta_\mathsf{C}} +  \omega_{\mathsf{I},1} \var_{\theta_{\mathsf{I},1}} }{\mu_1-\omega_{\mathsf{I},1}}   \frac{\pi e}{6} 2^{-2(R+R^*)}.
	\label{eq:perf_oracle_dsc_JR gaussian EC}
\end{align} 
Finally,
\begin{align}
D^{\JR}_{{x}_1|x_2}(R) &\ge D^{\IR}_{{x}_1}(R+R^*+R^{\mathsf{JR}}),
	\label{eq:RD for JR with rate gain}\\
\mbox{where }	R^{\mathsf{JR}}& = \frac{1}{2}\log_2 \left[\frac{1}{\omega_{\mathsf{I},1}}\frac{\mu_1-\omega_{\mathsf{I},1}}{\mu_1-1}\right]
\label{eq:rate gain IR_JR}
\end{align} 
and where $R^*$ has been defined in \eqref{eq:rate_gain}. 
Therefore, when the SI is available at the decoder, it helps reducing the rate by $R^*+R^{\mathsf{JR}}$ bits per measurement sample.
\end{theorem}

\noindent  \textbf{{Sketch of proof.}} An oracle is considered in order to derive lower bounds. More precisely, it is assumed that the sparsity support of $\x_1$ is known if independent reconstruction is performed and that also the support of the common component $\x_{\mathsf{C}}$ is known if joint reconstruction is performed. The exact distortion of the oracles are derived, from which a closed--form expression of the rate gains are given. See the complete proof in Appendix~\ref{annex:th:RD reconstruction}.
\hfill $\square$

As one would expect, when there is no innovation component ($\omega_{\mathsf{I},j}\to 0$), the distortion of the oracle is zero and the rate gain $R^{\mathsf{JR}}$ is largest (tends to infinity). On the contrary, when there is no common component ($\omega_{\mathsf{I},j} \to 1$), $R^{\mathsf{JR}}$  tends to zero. Moreover, even if  the SI could be exploited also to enhance the quality of the dequantization of the unknown source, the gain due to Joint Dequantization becomes negligible in the high--rate region \cite{coluccia2011lossy}. For this reason we neglected Joint Dequantization in this analysis.

\section{Numerical Results}\label{sec:num_res}

In this section, we validate by simulations the results obtained in the previous sections, comparing  numerical results to the derived equations. In particular, first we validate the RD functions of the measurement vector, both in the single--source and in the distributed case, hence validating at the same time the rate gain $R^*$. Then, we validate the RD functions of independent and joint reconstruction (hence the rate gain $R^\mathsf{JR}$). Finally, we compare the results obtained using the oracle estimator with the results obtained using practical reconstruction algorithms, showing that the rate gains hold also in a practical scenario. For each test, \Ps\ is the DCT matrix, each non zero component of $\thet$ is drawn from a normal distribution, and $\var_{\Ph} = 1$.

\begin{figure}
\centering
\subfloat[$N = 512$, $K_\mathsf{C} = K_{\mathsf{I},j} = 8$, $\var_{\theta_\mathsf{C}} = 1$, $\var_{\theta_{\mathsf{I},j}}=10^{-2}$, $M=128$]{\includegraphics[width=0.5\textwidth]{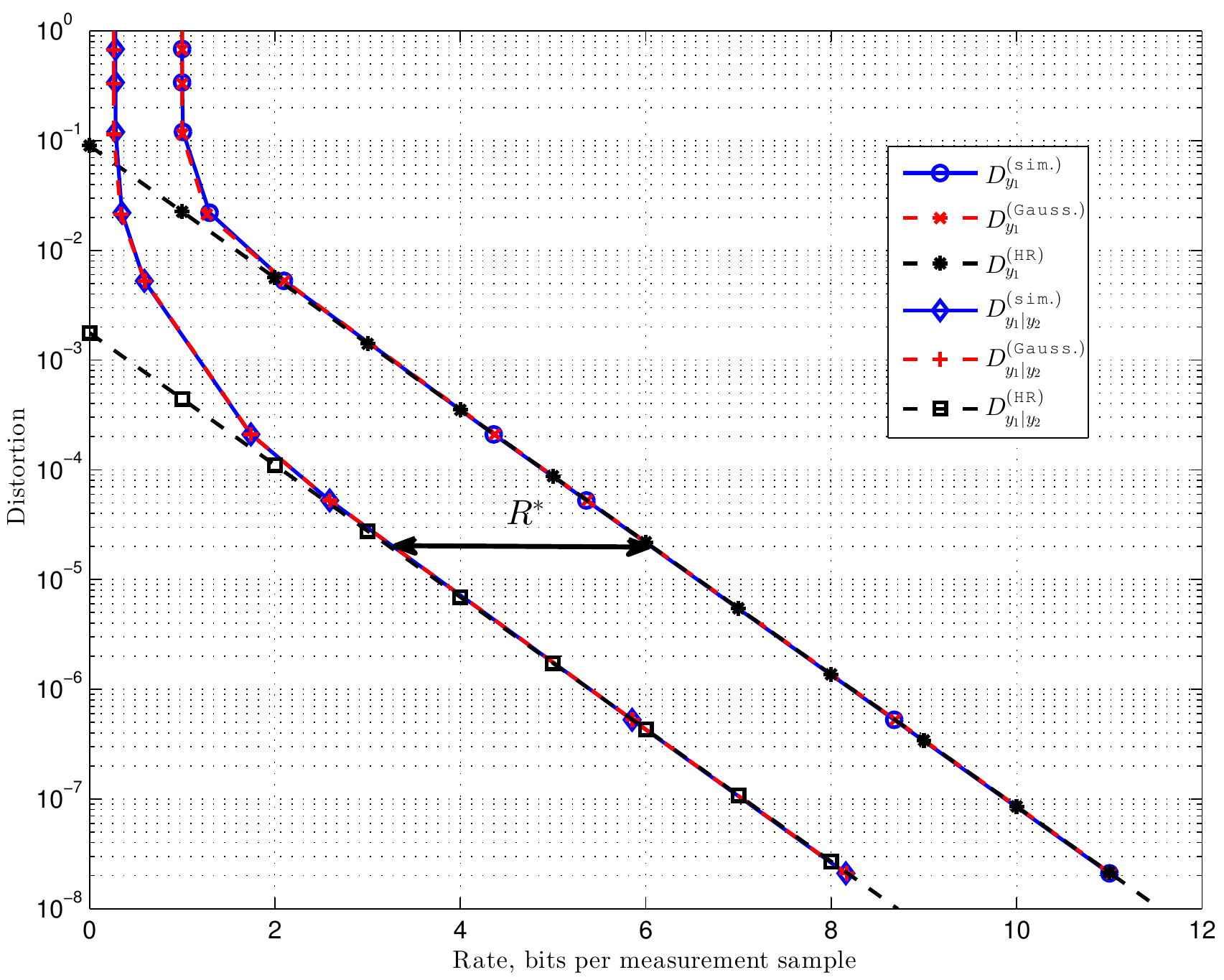}\label{fig:meas_1}}\\
\subfloat[$N = 1024$, $K_\mathsf{C} = 16$, $K_{\mathsf{I},j} = 8$, $\var_{\theta_\mathsf{C}} = \var_{\theta_{\mathsf{I},j}}= 1$, $M=256$]{\includegraphics[width=0.5\textwidth]{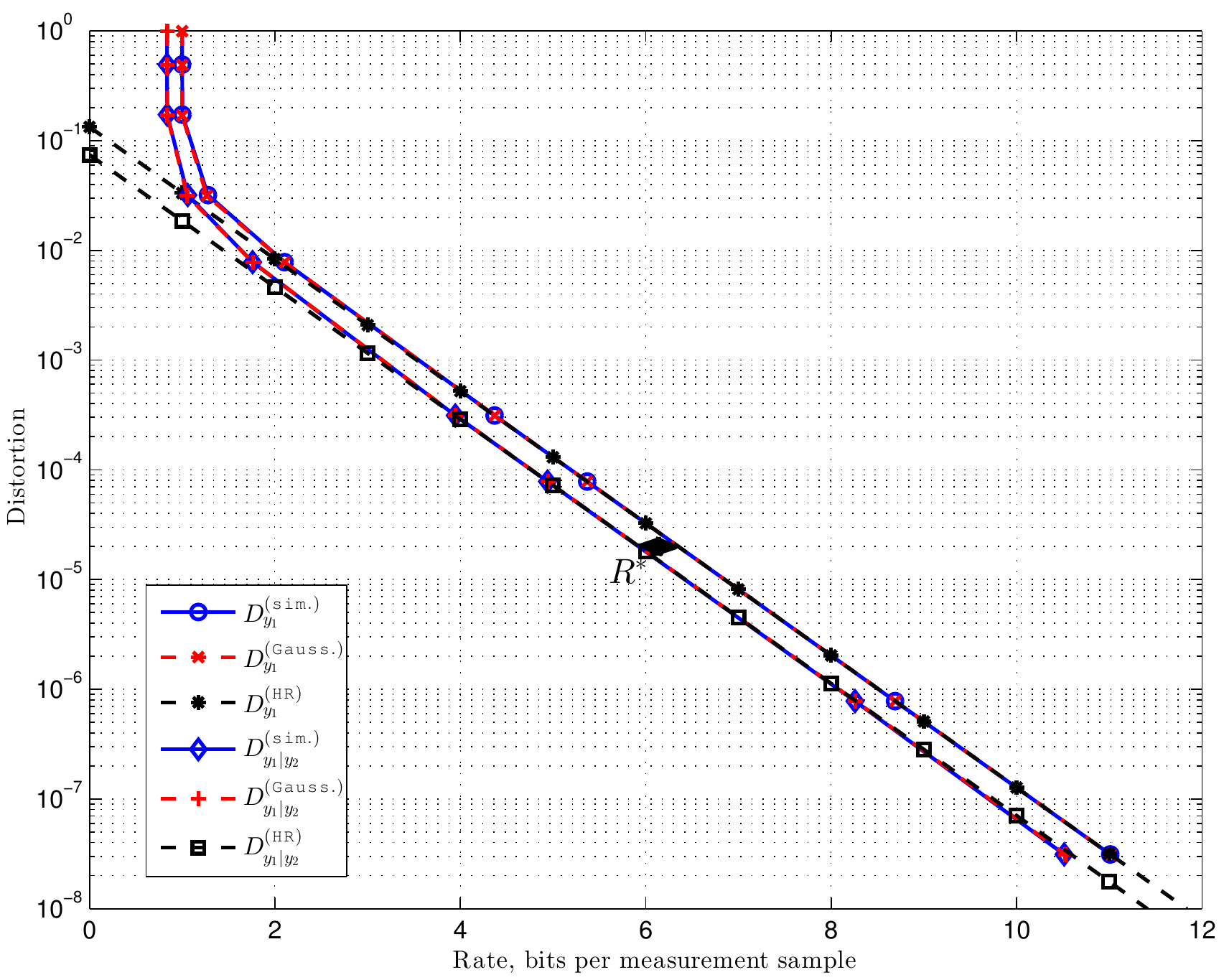}\label{fig:meas_2}}

\caption{Simulated vs. theoretical Rate--Distortion functions of \emph{measurement vectors} for two different sets of signal and system parameters. Single--source and distributed cases.}

\label{fig:rd_y_M128_espo-1}
\end{figure}

First, we test the validity of the RD functions derived for the measurements. Fig.~\ref{fig:rd_y_M128_espo-1} plots the quantization distortion of $\y_1$, \emph{i.e.}, $\mean{\frac{1}{M}\ltwonorm{\y_1-\y_{q,1}}^2}$ versus the rate $R$, measured in bits per measurement sample (bpms). The distortion has been averaged over $10^4$ trials, and for each trial different realizations of the sources, the sensing matrix and the noise have been drawn. Fig.~\ref{fig:rd_y_M128_espo-1} shows two subfigures corresponding to different sets of signal and system parameters (signal length $N$, sparsity of the common component $K_\mathsf{C}$ and of the innovation component $K_{\mathsf{I},j}$, variance of the nonzero components $\var_{\theta_\mathsf{C}}$ and $\var_{\theta_{\mathsf{I},j}}$, respectively, and length of the measurement vector $M$). Each subfigure shows two families of curves, corresponding to the cases in which $\y_2$ is (respectively, is not) used as SI. Each family is composed by 3 curves. 
$i)$ The curve labeled as \textsf{(HR)} -- standing for \emph{high rate} -- is the asymptote of the operational RD curves \eqref{eq:rd_ec_y1} (or \eqref{eq:rd_ec_y1y2}).
$ii)$ The curve labeled as \textsf{(Gauss.)} corresponds to the distortion of a synthetic correlated Gaussian source pair with covariance matrix as in \eqref{eq:inter-correlation-matrix12}, where $\var_{\y_j}$ is defined in \eqref{eq:var_y1} and $\rho_{{\y_1}{\y_2}}$ in \eqref{eq:rho_y1y2}, and quantized with a uniform scalar quantizer.
The rate is computed as the \emph{symbol} entropy of the samples $\y_{q,1}$ (the conditional \emph{symbol} entropy of $\y_{q,1}$ given $\y_{2}$), quantized with a uniform scalar quantizer. Entropy and conditional entropy have been evaluated computing the number of symbol occurrences over vectors of length $10^8$.
$iii)$ The curves labeled as \textsf{(sim.)} are the simulated RD for a measurement vector pair obtained generating $\x_1$ and $\x_2$ according to Def.~\ref{def:jsm1}, measuring them with the same \Ph\ to obtain $\y_1$ and $\y_2$ and quantizing $\y_1$ and $\y_2$ with a uniform scalar quantizer.

First, we notice that the \textsf{(HR)} equation perfectly matches the simulated curves when $R>2$, showing that the high--rate regime occurs for relative small values of $R$. Then, it can be noticed that \textsf{(Gauss.)} curves perfectly overlap the \textsf{(sim.)} ones, validating both equations \eqref{eq:var_y1} and \eqref{eq:rho_y1y2} and showing that the convergence to the Gaussian case occurs for low values of $N$, $M$, $K_C$, $K_{\mathsf{I},j}$. %In Fig.~\ref{fig:rd_y_M128_espo-1}, $R^* = 2.83$~bpms.

\begin{figure}
\centering
\subfloat[$N = 512$, $K_\mathsf{C} = K_{\mathsf{I},j} = 8$, $\var_{\theta_\mathsf{C}} = 1$, $\var_{\theta_{\mathsf{I},j}}=10^{-2}$,  $M=128$]{\includegraphics[width=0.5\textwidth]{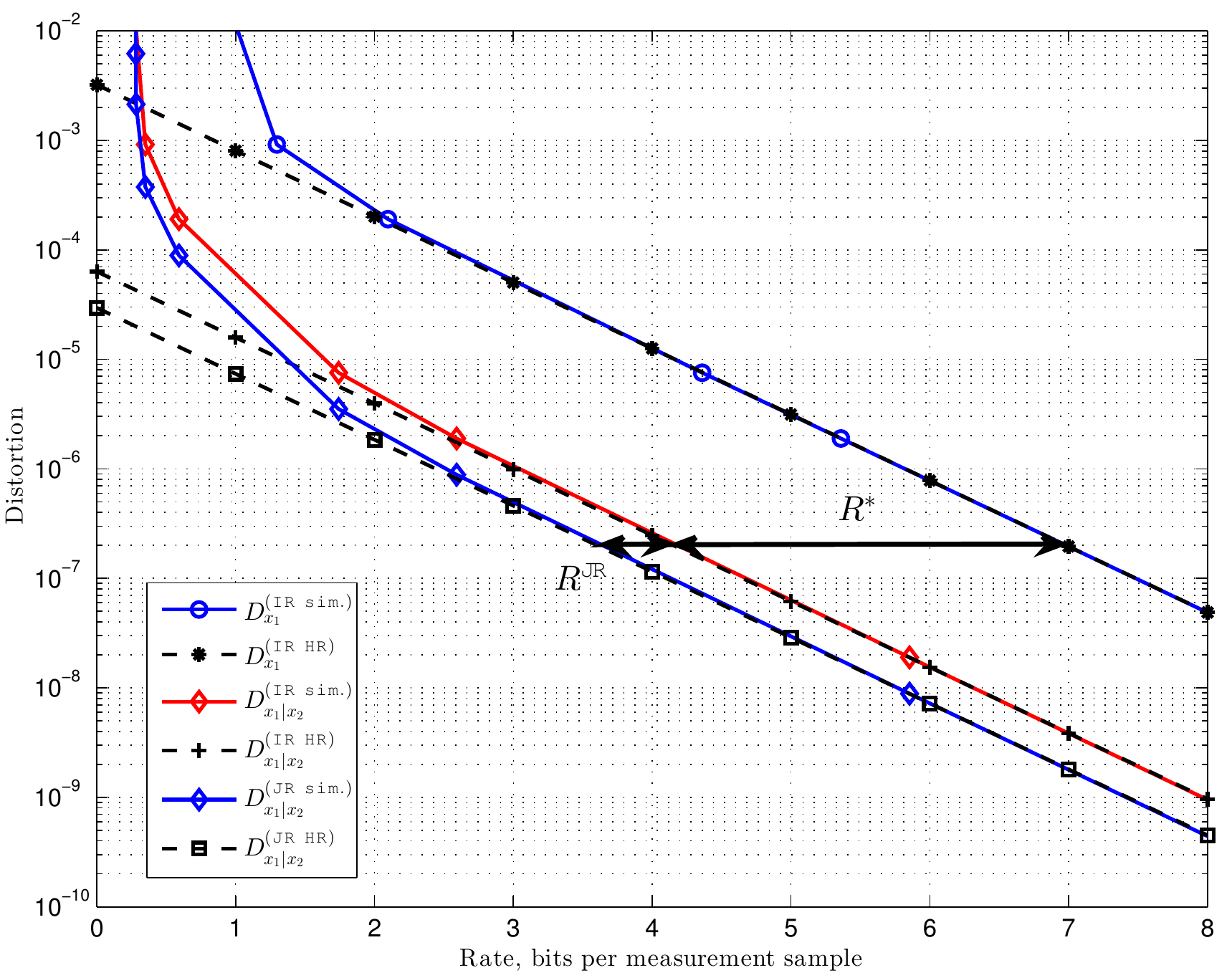}\label{fig:oracle_1}}\\
\subfloat[$N = 1024$, $K_\mathsf{C} = 16$, $K_{\mathsf{I},j} = 8$, $\var_{\theta_\mathsf{C}} = \var_{\theta_{\mathsf{I},j}}= 1$,  $M=256$]{\includegraphics[width=0.5\textwidth]{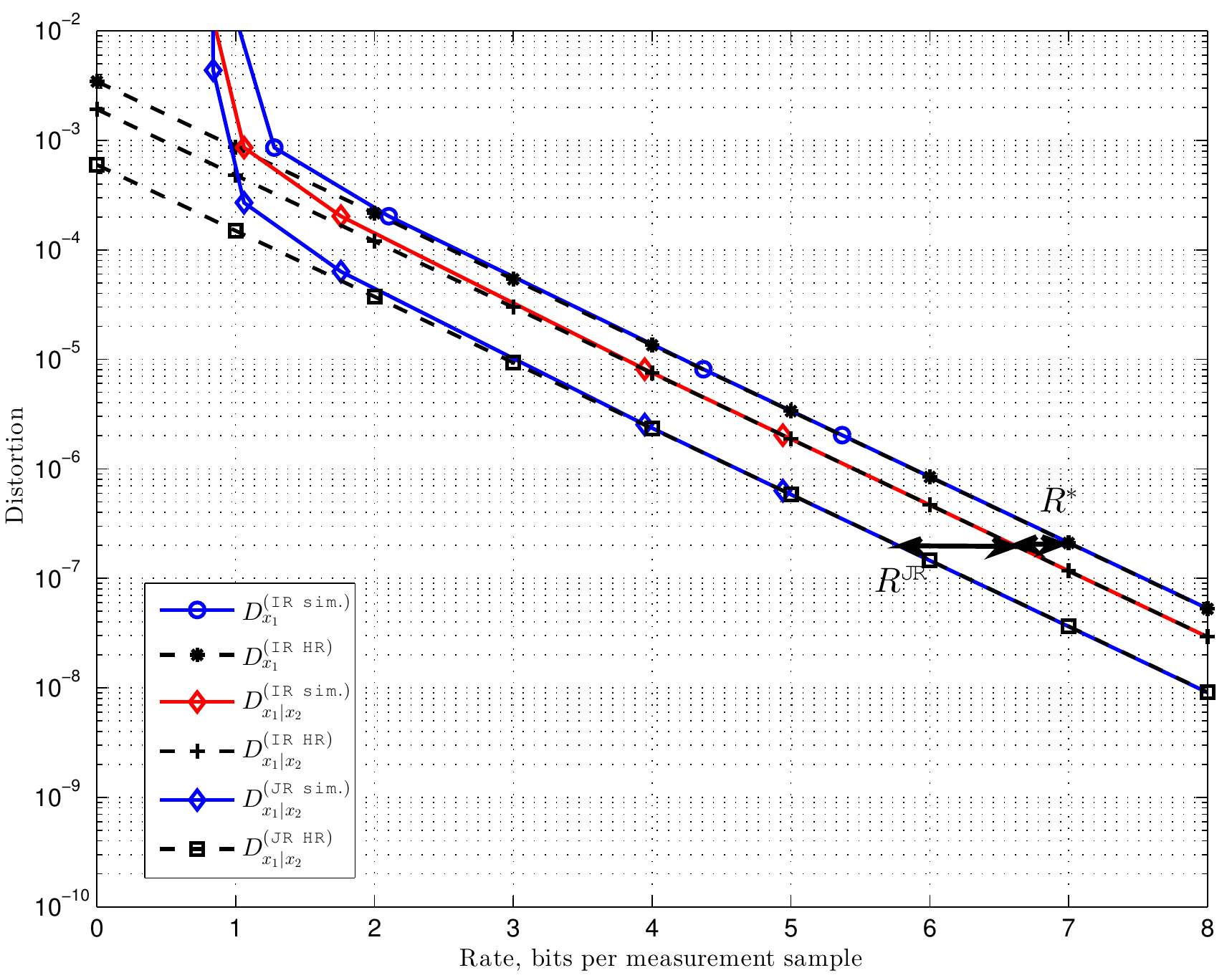}\label{fig:oracle_2}}
\caption{Simulated vs. theoretical Rate--Distortion functions of the \emph{oracle reconstruction} for two different sets of signal and system parameters. Single--source, distributed and joint reconstruction cases.}
\label{fig:rd_xrec_M128_espo-1}
\end{figure}

After validating the RD functions derived for the measurements, we test the RD functions for the oracle reconstruction. Fig.~\ref{fig:rd_xrec_M128_espo-1} depicts the performance of the complete DCS scheme, in terms of reconstruction error, \emph{i.e.}, $\mean{\frac{1}{N}\ltwonorm{\widehat{\x}_1-\x_{1}}^2}$ versus the rate per measurement sample $R$. The figure shows two subfigures corresponding to different sets of signal and system parameters ($N$,  $K_\mathsf{C}$, $K_{\mathsf{I},j}$, $\var_{\theta_\mathsf{C}}$, $\var_{\theta_{\mathsf{I},j}}$, $M$). Each subfigure shows three pairwise comparisons. First, it compares the RHS of equation \eqref{eq:perf_oracle_uncond gaussian EC} (\textsf{IR HR} -- standing for \emph{independent reconstruction high rate}) with the oracle reconstruction distortion from $\y_{q,1}$ vs. the \emph{symbol} entropy of the samples $\y_{q,1}$ (\textsf{IR sim.} -- standing for \emph{independent reconstruction simulated}), obtaining a match for $R>2$. Second, it compares the RHS of equation \eqref{eq:perf_oracle_dsc_IR gaussian EC} (\textsf{IR HR}) with the oracle reconstruction distortion from $\y_{q,1}$ vs. the conditional \emph{symbol} entropy of $\y_{q,1}$ given $\y_{2}$ (\textsf{IR sim.}), obtaining a match for $R>2.5$ and validating once more the evaluation of the rate gain $R^*$ due to the SWC. Third, it compares the RHS of equation \eqref{eq:perf_oracle_dsc_JR gaussian EC} (\textsf{JR HR} -- standing for \emph{joint reconstruction high rate}) with the ideal (knowing the sparsity support of the common component) oracle Joint Reconstruction distortion from  $\y_{q,1}$ and $\y_2$  vs. the conditional \emph{symbol} entropy of $\y_{q,1}$ given $\y_{2}$ (\textsf{JR sim.} -- standing for \emph{joint reconstruction simulated}), obtaining a match for $R>3$, validating the expression of the Rate Gain due to Joint Reconstruction given in \eqref{eq:rate gain IR_JR}. 

\begin{figure}
\centering
\includegraphics[width=0.5\textwidth]{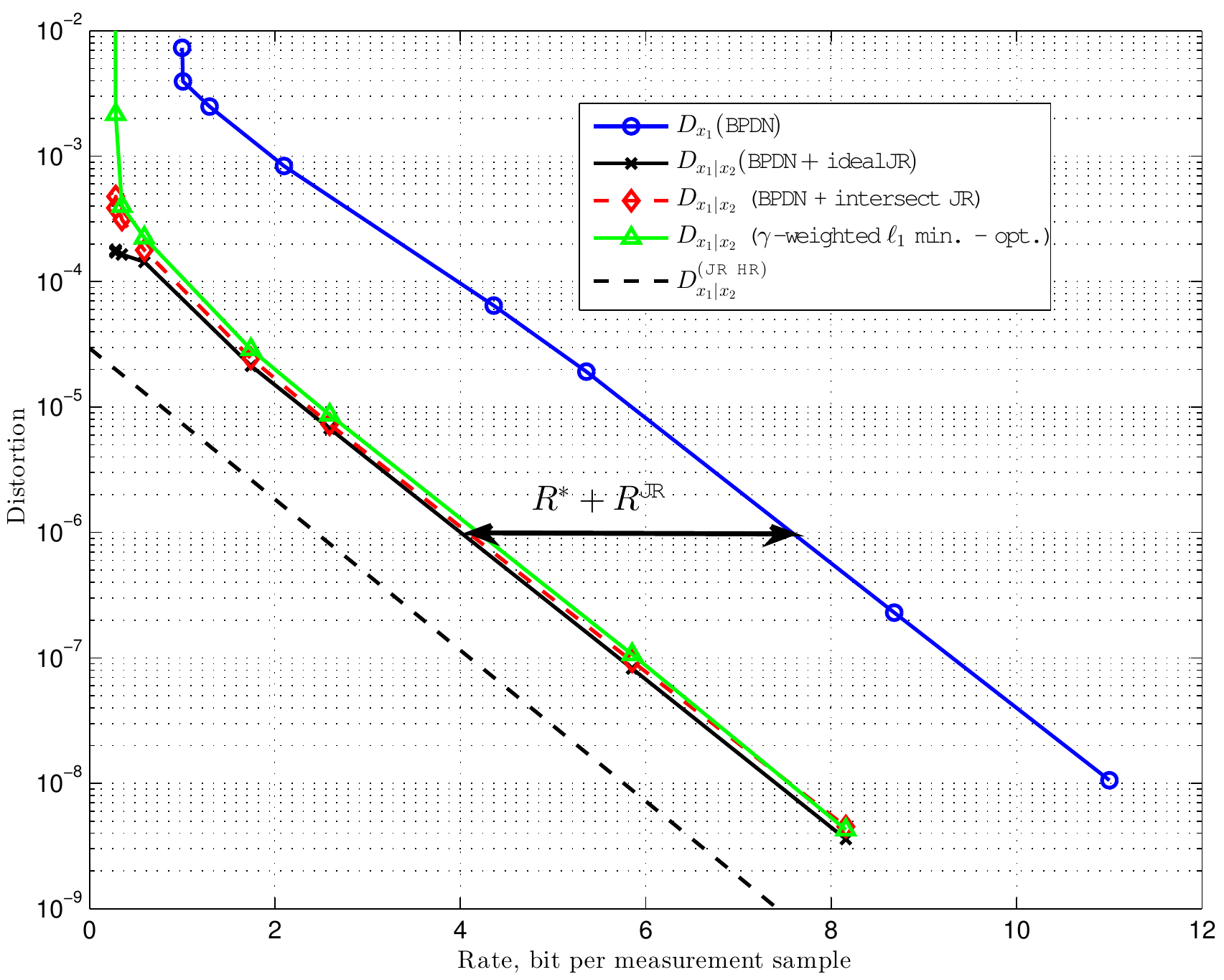}
\caption{Practical Reconstruction Rate--Distortion function. $N = 512$, $K_\mathsf{C} = K_{\mathsf{I},j} = 8$, $\var_{\Ph} = 1$, $\var_{\theta_\mathsf{C}} = 1$, $\var_{\theta_{\mathsf{I},j}}=10^{-2}$ and $M=128$}
\label{fig:rd_bpdn_M128_espo-1}
\end{figure}

Finally, we report in Fig.~\ref{fig:rd_bpdn_M128_espo-1} the performance of practical reconstruction algorithms solving the optimization problem \eqref{eq:CS_recovery_relaxed}. The curve labelled as \textsf{(BPDN)} reports the RD function of the independent reconstruction. The curve labelled as \textsf{(BPDN + ideal JR)} reports the RD function of the Joint Reconstruction when the sparsity support of the common component is known at the decoder. The curve labelled as \textsf{(BPDN + Intersect JR)}, instead, shows the RD performance of a Joint Reconstruction scheme in which the sparsity support of the common component is not known \emph{a priori}, but is estimated from the measurements $\y_1$ and $\y_2$ using the JR Algorithm 1 in \cite{coluccia2011lossy} (Intersect JR). The principle behind the Intersect JR algorithm is that the sparsity support of the common component is obtained as the intersection between the estimated sparsity supports of the information source and the SI. Comparing \textsf{(BPDN + ideal JR)} with the oracle performance curve \textsf{(JR HR)}, it can be noticed that, apart from a penalty due to the missing knowledge of the sparsity support, the slope of the RD curve is the same as in the ideal oracle case. Moreover, Fig.~\ref{fig:rd_bpdn_M128_espo-1} shows that the practical JR Algorithm 1 in \cite{coluccia2011lossy} performs very close to ideal JR algorithm. Note that a fully practical scheme will use a real SWC encoder/decoder. In \cite{coluccia2011lossy} we showed that a system implementing a real SWC encoder/decoder performs very close to the lower bound represented by the entropy and conditional entropy. Finally, in Fig.~\ref{fig:rd_bpdn_M128_espo-1} it can be seen that $R^*+R^\mathsf{JR} = 3.49$~bpms, which roughly corresponds to the sum of $R^* = 2.83$~bpms in Figs.~\ref{fig:meas_1} and \ref{fig:oracle_1} and $R^\mathsf{JR} = 0.55$~bpms in Fig.~\ref{fig:oracle_1},  proving that the Rate Gains derived for the Oracle receiver hold in a practical scenario, as well. Finally, Fig.~\ref{fig:rd_bpdn_M128_espo-1} plots the performance of the $\gamma$-weighted $\ell_1$-norm minimization algorithm \cite[(12)]{baron2005distributed}, which for a fair comparison we optimized to take into account that the measurements of the SI are perfectly known whereas the measurements of the unknown source are subject to quantization noise. It can be noticed that the performance is slightly worse with respect to the Intersect JR algorithm. Plus, the complexity increase is significant since the $\gamma$-weighted $\ell_1$-norm minimization algorithm needs the solution of a problem of size $3N$, which means a compexity increase of 8 times since the complexity of basis pursuit minimization algorithms is cubic with the size of the problem.

\section{Conclusions}\label{sec:conclusions}

We have studied the best achievable performance in the RD sense over all single--source and DCS schemes, under the constraint of high--rate quantization. 
Closed form expressions of the RD curves have been derived in the asymptotic regime, and simulations have shown that the asymptote is reached for relatively small number of measurements ($M\!\!\simeq\!\!100$) and small rate ($R>2$ bits/measurement sample). The RD curve computation is based on the convergence of the measurement vector to the multidimensional standard normal distribution. This generalizes \cite[Theorem 2]{dai2011quantized} that derives the marginal distribution of the measurement samples when the observed signal is directly sparse. We have then derived a closed--form expression of the highest rate gain achieved exploiting all levels of source correlations at the receiver, along with a closed--form expression of the average performance of the oracle receiver, using novel results on random Wishart matrix theory.
Simulations showed that the scheme proposed in \cite{coluccia2011lossy} almost achieves this best rate gain, and that the only penalty is due to the missing knowledge of the sparsity support as in single--source compressed sensing.

\appendices

\section{Proof of Theorem \ref{theorem:y iid gaussian}}
\label{annex:lemma y iid gaussian}
The Gaussian random matrix $\Ph$ is orthogonal invariant \cite[Section 4.3]{Edelman2005rmt}, \cite[Ex. 2.4]{tulino2004rmt}. Therefore, $\mat{U}=\Ph\Ps$ is a random matrix with i.i.d. entries drawn from $\N(0,\var_{\Phi})$. 
Without loss of generality, we assume that the $K$ non zeros of $\thet$ are placed at the beginning of $\thet$.
To show that the infinite length measurement random variables are asymptotically independent and Gaussian, we consider a finite length vector $\y^{(m)}$ that contains the $m$ first components of $\y$, and show that, for each $m$, $\y^{(m)}$ converges to a Gaussian vector with diagonal covariance matrix, as $N$ grows.

Let $\underline Y$ and \Uuk denote the random vectors in $\Ri^m$ associated to the Gaussian measurement vector $\y^{(m)}$ and to the $k$-th column of the matrix $\mat{U}$, restricted to its $m$ first rows. Let $\Theta_k$ denote the random variable in $\Ri$ associated to the $k$-th component of the vector $\thet$.
We have 
\begin{align}
	\underline Y & = \frac{1}{\sqrt{M}} \sum_{k=1}^K \Uuk \Theta_k = \sqrt{\frac{K}{M}} \frac{1}{\sqrt{K}} \sum_{k=1}^K \Uuk \Theta_k
\end{align}
By definition, $\Uuk$ and  $\Theta_k$ are independent, and the sequences $\{\Uuk\}_k$ and  $\{\Theta_k\}_k$ are i.i.d. centered with covariance matrix $\var_{\Phi}\I$  and variance $\var_{\theta}$, respectively. Therefore, each vector $\Uuk \Theta_k$ is centered with covariance matrix $\var_{\Phi} \var_{\theta} \I$, where $\I$ is the $m\times m$ identity matrix.
From the multidimensional Central Limit theorem \cite[Theorem 7.18]{khoshnevisan2007}, $Y$ converges (in distribution) to a multidimensional Gaussian distribution with mean vector $\0$ and covariance matrix $\frac{1}{\mu}{\var_{\Phi} \var_{\theta}}\I$. Thus, the entries of the vector are independent. Letting $m$ grow to $\infty$, we have that the sequence of Gaussian measurements converges to a Gaussian i.i.d. sequence with mean $0$ and variance $\frac{1}{\mu}{\var_{\Phi} \var_{\theta}}$.
\hfill $\square$

%-----------------------------------------------------------------------------------------
\section{Proof of Theorem \ref{th:RD reconstruction non distributed}}
\label{annex:th:RD reconstruction non distributed}
% begin proof
We derive a lower bound on the achievable distortion by assuming that the sparsity support $\Omega$ of $\x$ is known at the decoder. Let $\mat{U}_{\Omega}$ be the submatrix of $\U$ obtained by keeping the columns of $\Ph\Ps$ indexed by $\Omega$, and let $\Omega^c$ denote the complementary set of indexes. The optimal reconstruction is then obtained by using the pseudo--inverse of $\mat{U}_{\Omega}$, denoted by $\U_{\Omega}^\dagger$:
\begin{align}\label{eq:rec_oracle}
\left\{\begin{array}{ll}
\widehat\thet_{\Omega} & =  
\sqrt{M} \mat{U}^\dagger_{\Omega} \y_q := \sqrt{M} \left(\trasp{\mat{U}_{\Omega}}\mat{U}_{\Omega}\right)^{-1}\trasp{\mat{U}_{\Omega}}\y_q \\
\widehat\thet_{\Omega^c} & = \0 
\end{array}\right.
\end{align}
and $\widehat\x = \Ps \widehat\thet$, where $\y_q$ is the quantized version of $\y$  i.e. $\y_q=\y+\e$, where $\e$ is the quantization noise of variance $\sigma^2_e$. Note that in \eqref{eq:rec_oracle} the product by $\sqrt{M}$ is a consequence of Definition~\ref{def:measurement}.
\begin{align}
\mean{\ltwonorm{\widehat\x-\x}^2} &= \mean{\ltwonorm{\widehat\thet-\thet}^2} =  \mean{\ltwonorm{\widehat\thet_{\Omega}-\thet_{\Omega}}^2} 
\label{eq:rc 1}\\
&= M \mean{\ltwonorm{ \UpO \e}^2}
\label{eq:rc 2}\\
&= M \mean{\e^T \mean{(\UO\UO^T)^\dagger} \e }  
\label{eq:rc 3}
\end{align}
The first equality in \eqref{eq:rc 1} follows from the orthogonality of the matrix $\Ps$, whereas the second follows from the assumption that $\Omega$ is the true support of $\thet$. \eqref{eq:rc 2} derives from the definition of the pseudo-inverse, and \eqref{eq:rc 3} from $\UpOT \UpO = (\UO\UO^T)^\dagger$. Then, if $M>K+3$,
\begin{align}
\mean{\ltwonorm{\widehat\x-\x}^2} &= M \mean{\e^T \frac{K}{M (M-K-1)} \frac{1}{\sigma^2_\Phi} \I \  \e }
\label{eq:rc 4}\\
&= \frac{M K}{M-K-1}  \frac{\sigma_e^2}{\sigma^2_\Phi}
\label{eq:rc 5} \\
\sigma^2_{\hat x} &= \frac{K}{N} \frac{M}{M-K-1} \frac{\sigma_e^2}{\sigma^2_\Phi}\label{eq:rc 6}
\end{align}
where $\sigma^2_{\hat x}$ stands for the distortion of the oracle estimator at finite length $N$.
\eqref{eq:rc 4} comes from the fact that, since $M>K$, $\UO \UO^T$ is rank deficient and follows a singular $M$-variate Wishart distribution with $K$ degrees of freedom and scale matrix $\sigma^2_\Phi \I$ \cite{DiazGarcia2006}. Its pseudo-inverse follows a generalized inverse Wishart distribution, whose distribution is given in \cite{DiazGarcia2006} and mean in \cite[Theorem 2.1]{Cook2011}, under the assumption that $M>K+3$. Note that the distortion of the oracle only depends on the variance of the quantization noise and not on its covariance matrix. Therefore, our result holds even if the noise is correlated (for instance if vector quantization is used).  As a consequence, we can apply our result to any quantization algorithm.
Therefore as $N\to \infty$, if $\mu>1$
\begin{align}
D_{x}(R)& \ge D^{\text{oracle}}_{x}(R) = \frac{\gamma}{1-\mu} \frac{1}{\sigma^2_\Phi} D_{y}(R)
\label{eq:RD recons oracle}\\
&= \gamma \frac{\mu}{1-\mu} \sigma^2_\theta  2^{-2R}\nonumber
\end{align}
where the first equality is obtained by taking the limit  $N\to \infty$ of \eqref{eq:rc 6} and the second equality follows from Theorem~\ref{theorem:y iid gaussian}. Substituting the entropy-constrained RD function of the measurements \eqref{eq:rd_ec_y} in \eqref{eq:RD recons oracle} leads to the entropy-constrained reconstruction RD function \eqref{eq:EC RD recons x}.\hfill $\square$
% end proof

%-----------------------------------------------------------------------------------------
\section{Proof of Theorem \ref{prop:rd_dcs}}
\label{annex:prop:rd_dcs}
% begin proof \label{prop:rd_dcs}
We first consider non-overlapping sparse components, i.e. ${\thet}_\mathsf{C}$,
$\thet_{\mathsf{I},1}$ and $\thet_{\mathsf{I},2}$ have non overlapping sparsity supports. 
From Theorem~\ref{theorem:y iid gaussian}, $\mat{U}=\Ph\Ps$ is a random matrix with i.i.d. entries drawn from $\N(0,\var_{\Phi})$. 
Without loss of generality, we assume that all non zeros of $\thet$ are placed at the beginning of $\thet$, with
first the $K_{\mathsf{C}}$ common, then the $K_{\mathsf{I},1}$ and the $K_{\mathsf{I},2}$ individual components. 

We build the finite length $2m$ vector $\y^{(2m)}$ that contains the $m$ first components of $\y_1$ concatenated with the $m$ first components of $\y_2$, and let $N$ go to infinity.
Let $\underline Y$ denote the random vector in $\Ri^{2m}$ associated to the Gaussian measurement vector $\y^{(2m)}$ and let  $\Uuk$ denote the random vector in $\Ri^{m}$ associated to the $k$-th column of the matrix $U$, restricted to its $m$ first rows. Let $\Theta_k$ denote the random variable in $\Ri$ associated to the $k$-th component of the vector $\thet$.
By definition of the sparse vectors and their measurements,  we have
\begin{align}
\underline  Y & = \frac{1}{\sqrt{M}} 
\left(\begin{array}{ll}
 \displaystyle \sum_{k=1}^{K_{\mathsf{C}}}  \Uuk \Theta_k  +  
 \displaystyle  \!\!\!\!\! \sum_{k=K_{\mathsf{C}} + 1}^{K_{\mathsf{C}} + K_{\mathsf{I},1} }  \!\!\!\!\! \Uuk \Theta_k 
  & + \ \0 \\
 \displaystyle \sum_{k=1}^{K_{\mathsf{C}}}  \Uuk \Theta_k  + \0 & + 
 \displaystyle  \!\!\! \!\!\!\!\! \sum_{k=K_{\mathsf{C}} + K_{\mathsf{I},1} +1}^{K_{\mathsf{C}} + K_{\mathsf{I},1}+ K_{\mathsf{I},2}}  \!\!\! \!\!\!\!\!  \Uuk \Theta_k 
 \end{array}\right) \nonumber \\
& =  \sqrt{\frac{K_{\mathsf{C}}}{M}} \YuC + 
	\sqrt{\frac{K_{\mathsf{I},1}}{M}} \YuIone +
  \sqrt{\frac{K_{\mathsf{I},2}}{M}} \YuItwo
\label{eq:Y2m}
\end{align}
where $\0$ stands for the all zero vector of length $m$ and \YuC, \YuIone and \YuItwo\ are defined as
\begin{align}
\YuC & := \frac{1}{\sqrt{K_{\mathsf{C}}}}\left(\begin{array}{c}\displaystyle\sum_{k=1}^{K_{\mathsf{C}}}  \Uuk \Theta_k\\
\displaystyle\sum_{k=1}^{K_{\mathsf{C}}}  \Uuk \Theta_k
\end{array}\right)\nonumber\\
\YuIone & := \frac{1}{\sqrt{K_{\mathsf{I},1}}}\left(\begin{array}{c}\displaystyle\sum_{k=K_{\mathsf{C}} + 1}^{K_{\mathsf{C}} + K_{\mathsf{I},1}}  \Uuk \Theta_k\\
\0
\end{array}\right)\nonumber\\
\YuItwo & := \frac{1}{\sqrt{K_{\mathsf{I},2}}}\left(\begin{array}{c}\0\\
\displaystyle\sum_{k=K_{\mathsf{C}} + K_{\mathsf{I},1} +1}^{K_{\mathsf{C}} + K_{\mathsf{I},1}+ K_{\mathsf{I},2}}  \Uuk \Theta_k
\end{array}\right)\nonumber
\end{align}

Each vector $\Uuk \Theta_k$ is centered with covariance matrix
\begin{align*}
\var_{\Phi} \var_{\theta_\mathsf{C}} \I\quad & \text{if} & 1\le k \le K_{\mathsf{C}}\\
\var_{\Phi} \var_{\theta_{\mathsf{I},1}} \I\quad & \text{if} & K_{\mathsf{C}}+1 \le k\le K_{\mathsf{C}} + K_{\mathsf{I},1}\\
\var_{\Phi} \var_{\theta_{\mathsf{I},2}} \I\quad & \text{if} &  K_{\mathsf{C}}+ K_{\mathsf{I},1}+1 \le k \le K_{\mathsf{C}} + K_{\mathsf{I},1} + K_{\mathsf{I},2}~,
\end{align*}
where $\I$ is the $m\times m$ identity matrix.

From the multidimensional CLT \cite[Theorem 7.18]{khoshnevisan2007}, \YuC	\ converges (in distribution) to a multidimensional Gaussian distribution with mean vector $\0$ and covariance matrix $\var_{\Phi} \var_{\theta_\mathsf{C}} \bigl(\begin{smallmatrix} \I&\I\\ \I&\I \end{smallmatrix} \bigr)$ in the large system regime. 
Similarly,  in the large system regime, \YuIone\ and \YuItwo\ converge (in distribution) to a multidimensional Gaussian distribution with mean vector $\0$ and covariance matrix $ \var_{\Phi} \var_{\theta_{\mathsf{I},1}} \bigl(\begin{smallmatrix} \I&\0\\ \0&\0 \end{smallmatrix} \bigr)$, $ \var_{\Phi} \var_{\theta_{\mathsf{I},2}} \bigl(\begin{smallmatrix} \0&\0\\ \0&\I \end{smallmatrix} \bigr)$ respectively, where $\0$ stands for the all zero $m\times m$ matrix\footnote{For the sake of brevity and when it is clear from the context, we use $\0$ to denote either an $m$ length vector or an  all zero $m\times m$ matrix.}. Moreover,  \YuC, \YuIone\ and \YuItwo\ are independent, since the elements of each sum are all different in \eqref{eq:Y2m}. Therefore, $\underline Y$ converges to a Gaussian centered vector with covariance matrix
\begin{equation}
\var_{\Phi}
\begin{pmatrix}
\left( \frac{\omega_{\mathsf{C},1}}{\mu_1}    \var_{\theta_\mathsf{C}} +  \frac{\omega_{\mathsf{I},1}}{\mu_1} \var_{\theta_{\mathsf{I},1}}  \right) \I&  \sqrt{\frac{\omega_{\mathsf{C},1}}{\mu_1}\frac{\omega_{\mathsf{C},2}}{\mu_2}} \var_{\theta_\mathsf{C}} \I\\ 
 \sqrt{\frac{\omega_{\mathsf{C},1}}{\mu_1}\frac{\omega_{\mathsf{C},2}}{\mu_2}} \var_{\theta_\mathsf{C}} \I& \left( \frac{\omega_{\mathsf{C},2}}{\mu_2} \var_{\theta_\mathsf{C}} +  \frac{\omega_{\mathsf{I},2}}{\mu_2} \var_{\theta_{\mathsf{I},2}}  \right) \I
\end{pmatrix}~,
\label{eq:cov matrix Y2m}
\end{equation}
since $\frac{\omega_{\mathsf{C},1}}{\mu_1} = \frac{\omega_{\mathsf{C},2}}{\mu_2}$. In the case of overlapping supports, each entry of the vector $\thet$ can either contain the contribution of only one component (common, or any innovation), or any combination of at least two  components. Therefore, the support of the non zero components of $\thet$ and therefore the random vector $\underline Y$, can be decomposed into 7 terms, that correspond to the number of subsets of a set of 3 elements (excluding the empty set). Note that each term is a sum of i.i.d. random vectors such that the multidimensional CLT still applies and \eqref{eq:cov matrix Y2m} holds. 

Letting $m$ grow to $\infty$, we have that, in the large system regime,
$(Y_{1},Y_{2})$ converges to an i.i.d. Gaussian sequence with covariance matrix
\begin{align}
\mat{K}_{12} & = \left(\begin{array}{cc}
\var_{y_1} & \rho_{12} \sigma_{y_1}\sigma_{y_2} \\
\rho_{12} \sigma_{y_1}\sigma_{y_2} & \var_{y_2}
\end{array}\right),\ \text{where}\nonumber\\ 
\sigma^2_{y_j} & = \frac{\var_{\Phi}}{\mu_j} \left[  \omega_{\mathsf{C},j} \var_{\theta_\mathsf{C}} +  \omega_{\mathsf{I},j} \var_{\theta_{\mathsf{I},j}} \right]\ \text{and} \nonumber\\
\rho_{12} &= \left[
\Big(1 + \frac{ \omega_{\mathsf{I},1}}{ \omega_{\mathsf{C},1}}\frac{\var_{\theta_{\mathsf{I},1}}}{\var_{\theta_\mathsf{C}}}\Big)
\Big(1 + \frac{ \omega_{\mathsf{I},2}}{ \omega_{\mathsf{C},2}}\frac{\var_{\theta_{\mathsf{I},2}}}{\var_{\theta_\mathsf{C}}}\Big)
\right]^{-\frac{1}{2}}~,
 \nonumber
\end{align}
since $\frac{\omega_{\mathsf{C},1}}{\mu_1} = \frac{\omega_{\mathsf{C},2}}{\mu_2}$.
Therefore, the RD functions are given in \eqref{eq:RD ec X}, \eqref{eq:RD X}, and
\eqref{eq:RD_WZ_def}.
\hfill $\square$
% end proof

%-----------------------------------------------------------------------------------------
\section{Proof of Theorem \ref{th:RD reconstruction}}
\label{annex:th:RD reconstruction}

Let us first consider independent reconstruction. This means that the measurements of the SI are used at the SWC decoder only and not to improve the quality of the dequantization or the reconstruction stages.
We derive a lower bound on the achievable distortion by assuming that the sparsity support $\Omega_1$ of $\x_1$ is known at the decoder. This receiver is called the oracle and leads to a variance of estimation $\sigma^2_{\hat x_1}$:
\begin{align}
\sigma^2_{\hat x_1} &= \frac{K_1}{N} \frac{M}{M-K_1-1} \frac{\sigma_e^2}{\sigma^2_\Phi}
\end{align}
where the derivation is similar to the non distributed case (see Appendix~\ref{annex:th:RD reconstruction non distributed}). $\sigma_e^2$ is the quantization noise variance, i.e. $D_{y_1|y_2}(R)$ if the $\y_2$ is used as side information at the SWC decoder and $D_{y_1}(R)$ otherwise. This leads to \eqref{eq:perf_oracle_uncond}  and \eqref{eq:perf_oracle_dsc_IR}. Then, in the large system regime, the measurements are Gaussian and the RD functions of the measurements have a closed form expression, which is used to derive \eqref{eq:perf_oracle_uncond gaussian}, \eqref{eq:perf_oracle_dsc_IR gaussian}, \eqref{eq:perf_oracle_uncond gaussian EC}, and \eqref{eq:perf_oracle_dsc_IR gaussian EC}.

Finally, note that the gap between the oracle based lower bound and the true RD functions, in the IR case, is only due to the performance of the  algorithm chosen to reconstruct $\x_1$ from $\y_1$. The performance of a deterministic CS reconstruction algorithm depends only on the density $p_{x_1|y_{q,1}}$, which in the present Gaussian case is determined by the MSE $D_{y_1}(R)$. In the IR case, $D_{y_1}(R)$ depends on the presence of the quantization/dequantization step, since the source coding/decoding step is considered a zero--error stage. Hence, the presence or absence of a (distributed) source encoding and decoding stage does not alter $D_{y_1}(R)$ since it does not alter the output of the dequantizer. Conditioning on $y_2$ yields a rate shift $D_{y_1|y_2}(R) = D_{y_1}(R+R^*)$ that will thus be directly reflected in the CS reconstruction. Hence, it can be written that $D^{\IR}_{{x}_1}(R) = f(D_{y_1}(R))$ and $D^{\IR}_{x_1|y_2}(R) = f(D_{y_1|y_2}(R)) = f(D_{y_1}(R+R^*))$ for some $f(\cdot)$ depending on the specific reconstruction algorithm. This yields \eqref{eq:RD for IR with rate gain gaussian}.

As for the joint reconstruction, the SI is used to reduce the sparsity level of the unknown signal, and hence the performance of the reconstruction. More precisely, to give a lower bound we assume that the receiver perfectly knows the common component $\x_\mathsf{C}$ and $\Omega_{\mathsf{I},1}$ of the innovation component $\x_{\mathsf{I},1}$. Hence, it will use the former to estimate the measurements of the common component $\y_\mathsf{C}=\Ph\x_\mathsf{C}$ and then subtract them from $\y_1$. In this way, the vector to be reconstructed is the innovation component $\x_{\mathsf{I},1}$ only, which is sparser than $\x_1$ (given the same $N$ and $M$). Then, it will use the latter information to apply the oracle,  leading to a variance of estimation $\sigma^2_{\hat x_1}$:
\begin{align}
\sigma^2_{\hat x_1} &= \frac{K_{\mathsf{I},1}}{N} \frac{M}{M-K_{\mathsf{I},1}-1} \frac{\sigma_e^2}{\sigma^2_\Phi}
\end{align}
where $\sigma_e^2$ is the measurement distortion, when $\y_2$ is used as side information at the receiver. This leads to
(\ref{eq:perf_oracle_dsc_JR}-\ref{eq:perf_oracle_dsc_JR gaussian EC}). \hfill $\square$

\section*{Acknowledgment}
The authors would like to thank the Associate Editor and the anonymous Reviewers for their valuable suggestions that helped to improve the quality of the final version of this paper.

\ifCLASSOPTIONcaptionsoff
  \newpage
\fi



%\bibliographystyle{IEEEtran}
%\bibliographystyle{IEEEtran}
%\bibliography{IEEEabrv,discos_spt}

\begin{thebibliography}{10}
\providecommand{\url}[1]{#1}
\csname url@samestyle\endcsname
\providecommand{\newblock}{\relax}
\providecommand{\bibinfo}[2]{#2}
\providecommand{\BIBentrySTDinterwordspacing}{\spaceskip=0pt\relax}
\providecommand{\BIBentryALTinterwordstretchfactor}{4}
\providecommand{\BIBentryALTinterwordspacing}{\spaceskip=\fontdimen2\font plus
\BIBentryALTinterwordstretchfactor\fontdimen3\font minus
  \fontdimen4\font\relax}
\providecommand{\BIBforeignlanguage}[2]{{%
\expandafter\ifx\csname l@#1\endcsname\relax
\typeout{** WARNING: IEEEtran.bst: No hyphenation pattern has been}%
\typeout{** loaded for the language `#1'. Using the pattern for}%
\typeout{** the default language instead.}%
\else
\language=\csname l@#1\endcsname
\fi
#2}}
\providecommand{\BIBdecl}{\relax}
\BIBdecl

\bibitem{donoho2006cs}
D.~Donoho, ``{Compressed sensing},'' \emph{IEEE Trans. on Information Theory},
  vol.~52, no.~4, pp. 1289--1306, 2006.

\bibitem{candes2006nos}
E.~Cand{\`e}s and T.~Tao, ``{Near-Optimal Signal Recovery From Random
  Projections: Universal Encoding Strategies?}'' \emph{IEEE Transactions on
  Information Theory}, vol.~52, no.~12, pp. 5406--5425, 2006.

\bibitem{dai2011quantized}
W.~Dai and O.~Milenkovic, ``Information theoretical and algorithmic approaches
  to quantized compressive sensing,'' \emph{IEEE Trans. on Communications},
  vol.~59, pp. 1857--1866, 2011.

\bibitem{weidmann12RDsparse}
C.~Weidmann and M.~Vetterli, ``Rate distortion behavior of sparse sources,''
  \emph{IEEE Trans. on Information Theory}, vol.~58, no.~8, pp. 4969--4992,
  Aug. 2012.

\bibitem{fletcher2007rate}
A.~Fletcher, S.~Rangan, and V.~Goyal, ``{On the rate-distortion performance of
  compressed sensing},'' in \emph{IEEE Int. Conf. on Acoustics, Speech and
  Signal Processing (ICASSP)}, vol.~3.\hskip 1em plus 0.5em minus 0.4em\relax
  IEEE, 2007.

\bibitem{goyal2008compressive}
V.~Goyal, A.~Fletcher, and S.~Rangan, ``{Compressive sampling and lossy
  compression},'' \emph{IEEE Signal Processing Mag.}, vol.~25, no.~2, pp.
  48--56, 2008.

\bibitem{baron2005distributed}
D.~{Baron}, M.~F. {Duarte}, M.~B. {Wakin}, S.~{Sarvotham}, and R.~G.
  {Baraniuk}, ``{Distributed Compressive Sensing},'' \emph{ArXiv e-prints},
  Jan. 2009.

\bibitem{duarte2013distributed}
M.~Duarte, M.~Wakin, D.~Baron, S.~Sarvotham, and R.~Baraniuk, ``Measurement
  bounds for sparse signal ensembles via graphical models,'' \emph{IEEE
  Transactions on Information Theory}, vol.~59, no.~7, pp. 4280--4289, 2013.

\bibitem{coluccia2011lossy}
G.~Coluccia, E.~Magli, A.~Roumy, and V.~Toto-Zarasoa, ``Lossy compression of
  distributed sparse sources: a practical scheme,'' in \emph{European Signal
  Processing Conf. (EUSIPCO)}, August 2011.

\bibitem{liu2006slepian}
Z.~Liu, S.~Cheng, A.~D. Liveris, and Z.~Xiong, ``{Slepian-Wolf coded nested
  lattice quantization for Wyner-Ziv coding: High-rate performance analysis and
  code design},'' \emph{IEEE Transactions on Information Theory}, vol.~52,
  no.~10, pp. 4358--4379, 2006.

\bibitem{DBLP:journals/corr/abs-1104-4842}
M.~A. Davenport, J.~N. Laska, J.~R. Treichler, and R.~G. Baraniuk, ``The pros
  and cons of compressive sensing for wideband signal acquisition: Noise
  folding vs. dynamic range,'' \emph{CoRR}, vol. abs/1104.4842, 2011.

\bibitem{laska2012regime}
J.~N. Laska and R.~G. Baraniuk, ``Regime change: Bit-depth versus
  measurement-rate in compressive sensing,'' \emph{Signal Processing, IEEE
  Transactions on}, vol.~60, no.~7, pp. 3496--3505, 2012.

\bibitem{Cook2011}
R.~D. Cook and L.~Forzani, ``On the mean and variance of the generalized
  inverse of a singular wishart matrix,'' \emph{Electronic Journal of
  Statistics}, vol.~5, pp. 146--158, 2011.

\bibitem{slepian1973noiseless}
D.~Slepian and J.~Wolf, ``Noiseless coding of correlated information sources,''
  \emph{IEEE Trans. on Information Theory}, vol.~19, no.~4, pp. 471--480, 1973.

\bibitem{cover1975aproof}
T.~Cover, ``{A proof of the data compression theorem of Slepian and Wolf for
  ergodic sources (Corresp.)},'' \emph{IEEE Trans. on Information Theory},
  vol.~21, no.~2, pp. 226--228, Sep. 1975.

\bibitem{rebollo2006highrate}
D.~Rebollo-Monedero, S.~Rane, A.~Aaron, and B.~Girod, ``High-rate quantization
  and transform coding with side information at the decoder,'' \emph{Signal
  Process.}, vol.~86, no.~11, pp. 3160--3179, Nov. 2006.

\bibitem{wyner1978continuousiid}
A.~Wyner, ``The rate-distortion function for source coding with side
  information at the decoder-ii: General sources,'' \emph{Information and
  Control}, vol.~38, no.~1, pp. 60--80, 1978.

\bibitem{bassi2010wyner}
F.~Bassi, M.~Kieffer, and C.~Weidmann, ``{Wyner-Ziv coding with uncertain side
  information quality},'' in \emph{European Signal Processing Conf. (EUSIPCO)},
  2010.

\bibitem{eldar2012compressed}
Y.~Eldar and G.~Kutyniok, Eds., \emph{Compressed Sensing: Theory and
  Applications}.\hskip 1em plus 0.5em minus 0.4em\relax Cambridge University
  Press, 2012.

\bibitem{baraniuk2008spr}
R.~Baraniuk, M.~Davenport, R.~DeVore, and M.~Wakin, ``{A simple proof of the
  restricted isometry property for random matrices},'' \emph{Constructive
  Approximation}, vol.~28, no.~3, pp. 253--263, 2008.

\bibitem{khoshnevisan2007}
D.~Khoshnevisan, \emph{Probability}, ser. Graduate Studies in
  Mathematics.\hskip 1em plus 0.5em minus 0.4em\relax AMS, 2007.

\bibitem{candes2006ssr}
E.~Cand{\`e}s, J.~Romberg, and T.~Tao, ``{Stable signal recovery from
  incomplete and inaccurate measurements},'' \emph{Communications on Pure and
  Applied Mathematics}, vol.~59, no.~8, 2006.

\bibitem{candes2008restricted}
E.~Cand{\`e}s, ``The restricted isometry property and its implications for
  compressed sensing,'' \emph{Comptes Rendus Mathematiques}, vol. 346, no.
  9-10, pp. 589--592, 2008.

\bibitem{DiazGarcia2006}
J.~A. D{\'\i}az-Garc{\'\i}a and R.~Guti{\'e}rrez-J{\'a}imez, ``Distribution of
  the generalised inverse of a random matrix and its applications,''
  \emph{Journal of statistical planning and inference}, vol. 136, no.~1, pp.
  183--192, 2006.

\bibitem{Edelman2005rmt}
A.~Edelman and N.~R. Rao, ``Random matrix theory,'' \emph{Acta Numerica},
  vol.~14, pp. 233--297, 4 2005.

\bibitem{tulino2004rmt}
\BIBentryALTinterwordspacing
A.~M. Tulino and S.~Verd\'{u}, ``Random matrix theory and wireless
  communications,'' \emph{Foundations and Trends in Commun. Inf. Theory},
  vol.~1, no.~1, pp. 1--182, Jun. 2004. [Online]. Available:
  \url{http://dx.doi.org/10.1516/0100000001}
\BIBentrySTDinterwordspacing

\end{thebibliography}


\begin{IEEEbiography}[{\includegraphics[width=1in,height=1.25in,clip,keepaspectratio]{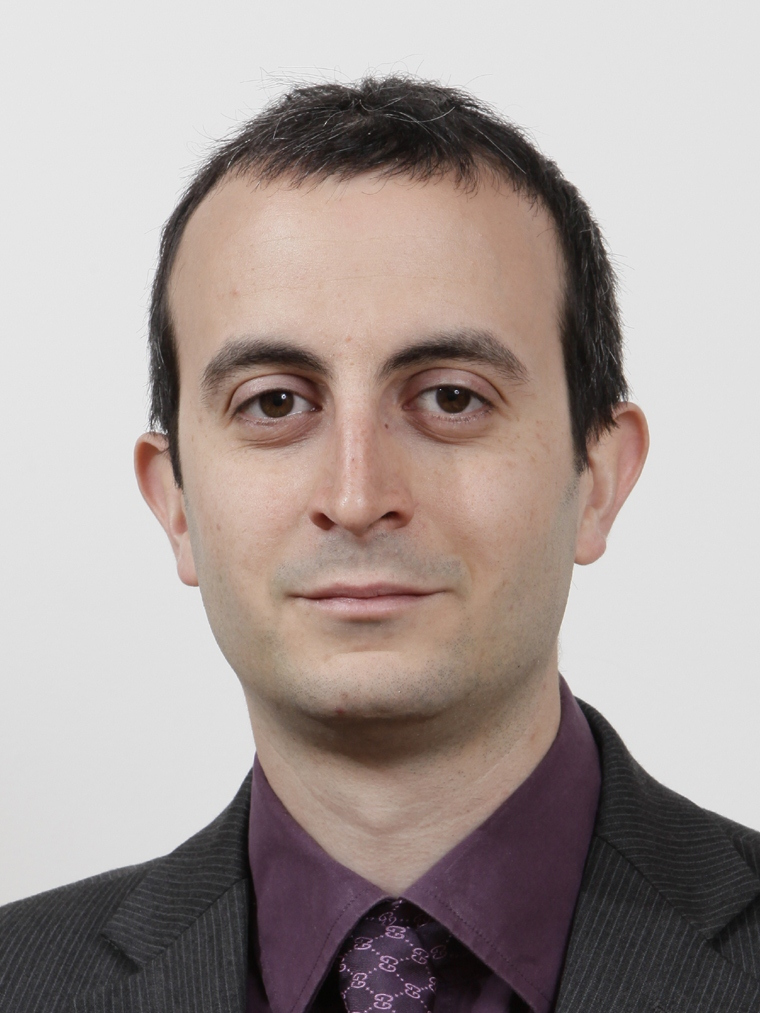}}]{Giulio Coluccia}
Giulio Coluccia received the bachelor in science (cum laude) in 2003 and the master in science (cum laude) in 2005 in Telecommunications Engineering, both from the Politecnico di Torino, Torino, Italy. He was a Ph.D. student within the Electronics Department of the Politecnico di Torino, Torino, Italy, under the supervision of prof. Giorgio Taricco. During his Ph.D. program, he visited the Telecommunications Research Center Vienna (ftw.), Vienna, Austria. He received the Ph.D. degree in Electronic and Communications engineering in 2009. His research activity included MIMO communications and space-time detection.

Currently, he is a Post Doctoral Researcher within the Image Processing Lab at Politecnico di Torino, leaded by Prof. Enrico Magli. His resarch is focused on Compressed Sensing, with particular interest in its application to Image Processing, Multidimensional Signals and to Distributed Source Coding and Wireless Sensor Netowrks. He is involved in the 5-year project entitled "CRISP - Towards compressive information processing systems" funded by the European Union.

In 2011 he received the Best Paper Award for Signal Processing at the GTTI (Gruppo Telecomunicazioni e Tecnologie dell'Informazione - Telecommunications and Information Technology Group) meeting.
\end{IEEEbiography}

\begin{IEEEbiography}[{\includegraphics[width=1in,height=1.25in,clip,keepaspectratio]{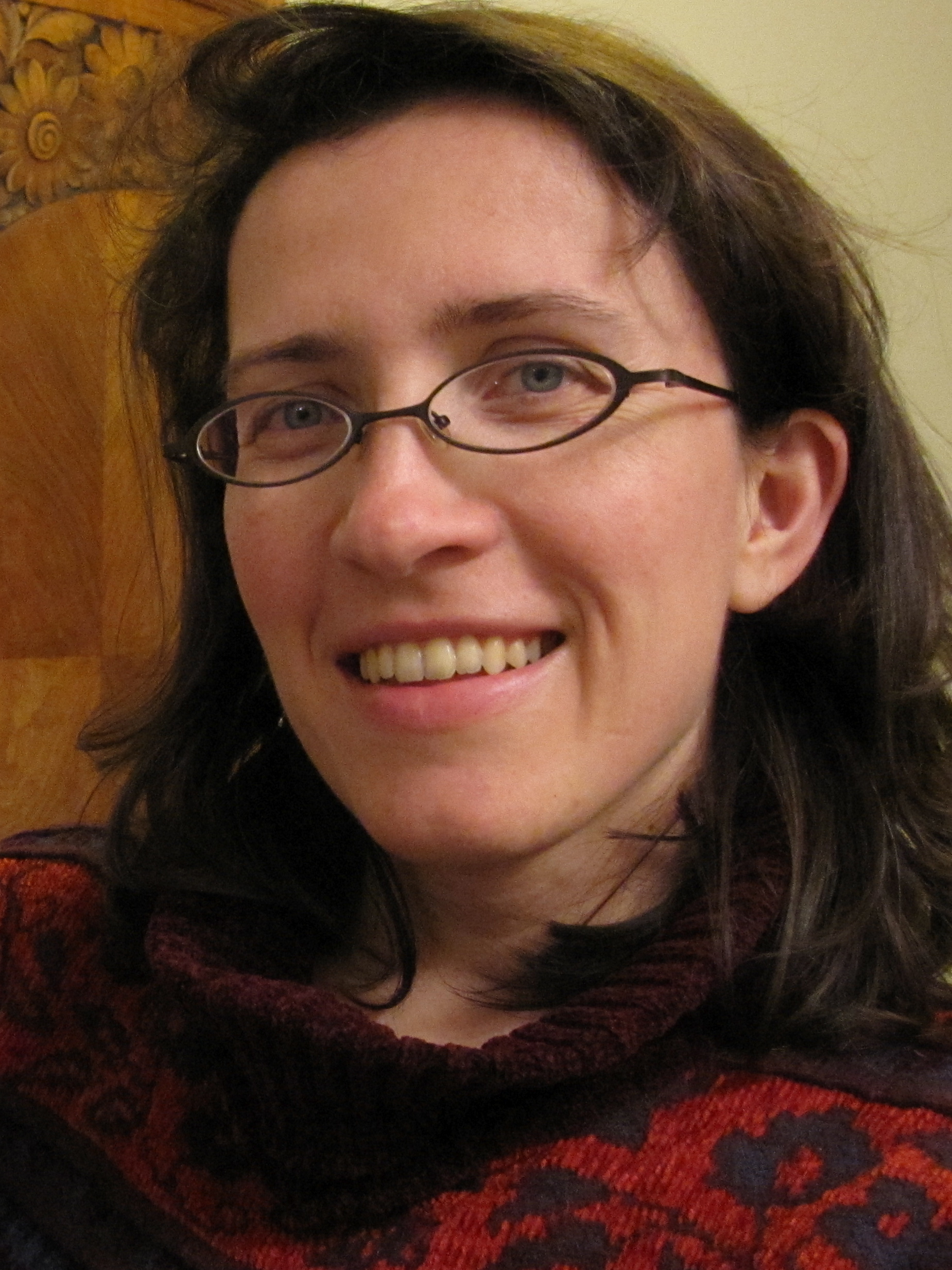}}]{Aline Roumy}
Aline Roumy received the Engineering degree from Ecole Nationale Superieure de l'Electronique et de ses Applications (ENSEA), Cergy, France in 1996, the Master degree in 1997 and the Ph.D. degree in 2000 from the University of Cergy-Pontoise, France. During 2000-2001, she was a research associate at Princeton University, Princeton, NJ. In 2001, she joined INRIA, Rennes, France as a research scientist. She has held visiting positions at Eurecom and Berkeley University.

Her current research and study interests include the area of statistical signal and image processing, coding theory and information theory. 

She has been a Technical Program Committee member and session chair at several international conferences, including ISIT, ICASSP, Eusipco. She is currently serving as a member of the French National University Council (CNU 61). She received the 2011 "Francesco Carassa" Best paper award.
\end{IEEEbiography}
\vfill
\newpage

\begin{IEEEbiography}[{\includegraphics[width=1in,height=1.25in,clip,keepaspectratio]{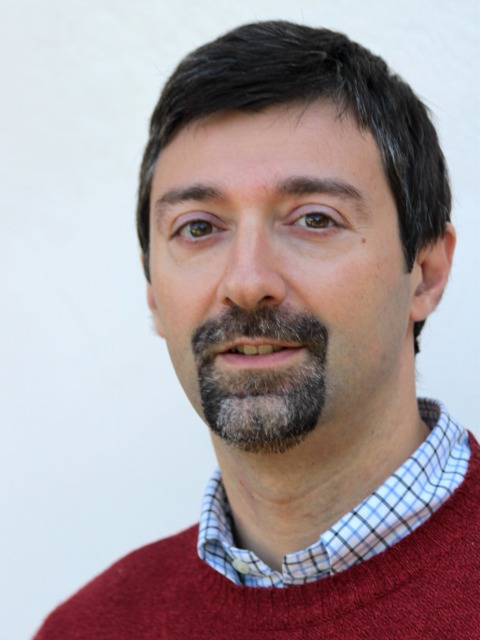}}]{Enrico Magli}
Enrico Magli received the Ph.D. degree in Electrical Engineering in 2001, from Politecnico di Torino, Italy. He is currently an Associate Professor at the same university, where he leads the Image Processing Lab. His research interests are in the field of multimedia signal processing and networking, compressive sensing, distributed source coding, image and video security, and compression of satellite images.

He is an associate editor of the IEEE Transactions on Circuits and Systems for Video Technology, and of the IEEE Transactions on Multimedia. He is a member of the Multimedia Signal Processing technical committee of the IEEE Signal Processing Society, and of the Multimedia Systems and Applications and the Visual Signal Processing and Communications technical committees of the IEEE Circuits and Systems Society. He has been co-editor of JPEG 2000 Part 11 - Wireless JPEG 2000. He is general co-chair of IEEE MMSP 2013, and has been TPC co-chair of ICME 2012, VCIP 2012, MMSP 2011 and IMAP 2007.

He has published about 40 papers in refereed international journals, 3 book chapters, and over 100 conference papers. He is a co-recipient of the IEEE Geoscience and Remote Sensing Society 2011 Transactions Prize Paper Award, and has received the 2010 Best Reviewer Award of IEEE Journal of Selected Topics in Applied Earth Observation and Remote Sensing.  He has received a 2010 Best Associate Editor Award of IEEE Transactions on Circuits and Systems for Video Technology. 
\end{IEEEbiography}

\vfill



%\vfill


\end{document}